\documentclass[journal]{IEEEtran}

\usepackage{graphicx}
\usepackage{amsmath}
\usepackage{amssymb}
\usepackage{afterpage}
\usepackage{verbatim}
\usepackage{psfrag}
\usepackage{algorithm}
\usepackage{lipsum}
\usepackage{color}
\usepackage{cite}
\usepackage{subfigure}
\usepackage{setspace}
\usepackage{epsfig}
\usepackage{epstopdf}
\usepackage{footmisc}
\usepackage{fmtcount}
\usepackage{stackrel}
\usepackage{abraces}
\usepackage{float}
\usepackage{soul}
\usepackage{bm}
\usepackage{url}
\usepackage{flushend}
\usepackage{enumerate}
\usepackage{color}
\usepackage{dblfloatfix}
\usepackage{tabularx,booktabs}
\usepackage{algorithmicx}

\begin{document}
%\doublespace

\title{Deep Learning in Industrial Internet of Things: Potentials, Challenges, and Emerging Applications}

\author{Ruhul~Amin~Khalil~\IEEEmembership{Student Member,~IEEE,} Nasir~Saeed~\IEEEmembership{Senior Member,~IEEE,} Yasaman Moradi Fard, Tareq Y. Al-Naffouri,~\IEEEmembership{Senior Member,~IEEE}, Mohamed-Slim Alouini,~\IEEEmembership{Fellow,~IEEE}
\thanks{Ruhul Amin Khalil is with the Department of Electrical Engineering,  University of Engineering and Technology, Peshawar 25120, Pakistan. email:ruhulamin@uetpeshawar.edu.pk}
\thanks{Nasir Saeed, Tareq Y. Al-Naffouri, and Mohamed-Slim Alouini are with Computer, Electrical, and Mathematical Sciences \& Engineering (CEMSE) Division, King Abdullah University of Science and Technology, Thuwal 23955, Makkah, Kingdom of Saudi Arabia. email: mr.nasir.saeed@ieee.org, }
\thanks{Yasaman Moradi Fard is with the Department of Biomedical Engineering, Faculty of Engineering, University of Isfahan, Isfahan, Iran. email: yasman.moradifard@gmail.com}
%\thanks{Manuscript received April 19, 2005; revised August 26, 2015.}
}

%\markboth{Journal of \LaTeX\ Class Files,~Vol.~14, No.~8, August~2015}
%{Shell \MakeLowercase{\textit{et al.}}: Bare Demo of IEEEtran.cls for IEEE Journals}

\maketitle{}
\begin{abstract}
The recent advancements in the Internet of Things (IoT) are giving rise to the proliferation of interconnected devices, enabling various smart applications. These enormous number of IoT devices generates a large capacity of data that further require intelligent data analysis and processing methods, such as Deep Learning (DL). Notably, the DL algorithms, when applied in the Industrial Internet of Things (IIoT), can enable various applications such as smart assembling, smart manufacturing, efficient networking, and accident detection-and-prevention. Therefore, motivated by these numerous applications; in this paper, we present the key potentials of DL in IIoT. First, we review various DL techniques, including convolutional neural networks, auto-encoders, and recurrent neural networks and there use in different industries. Then, we outline numerous use cases of DL for IIoT systems, including smart manufacturing, smart metering, smart agriculture, etc. Moreover, we categorize several research challenges regarding the effective design and appropriate implementation of DL-IIoT. Finally, we present several future research directions to inspire and motivate further research in this area.
\end{abstract}


\maketitle

\IEEEdisplaynotcompsoctitleabstractindextext
\IEEEpeerreviewmaketitle

\section{Introduction}
\begin{figure}[h!]
\begin{center}  
\includegraphics[width=0.735\columnwidth]{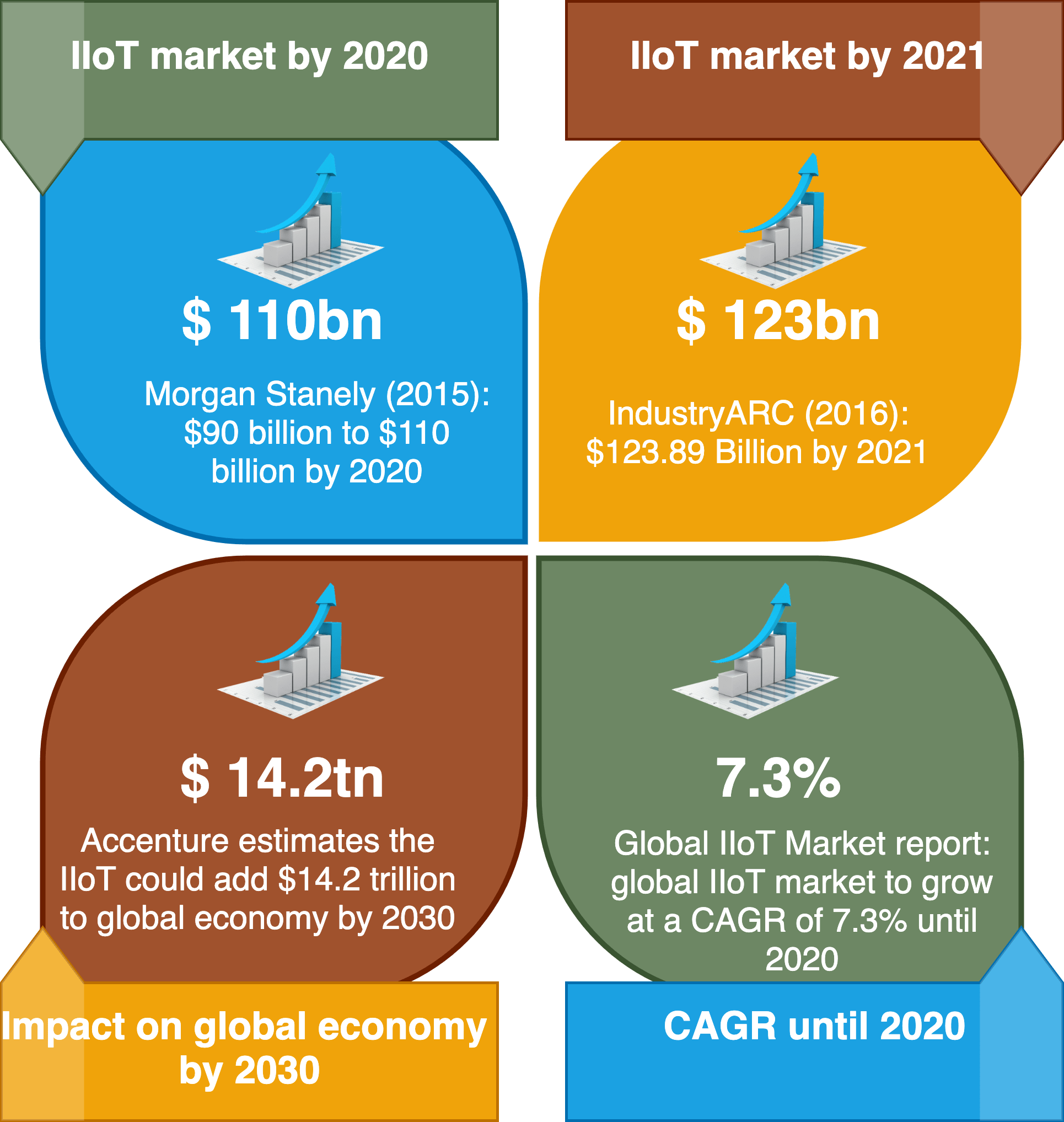}  
\caption{Size and market impact of IIoT .\label{sizemarket}}  
\end{center} 
\end{figure}

\begin{figure}[h!]
\begin{center}  
\includegraphics[width=0.835\columnwidth]{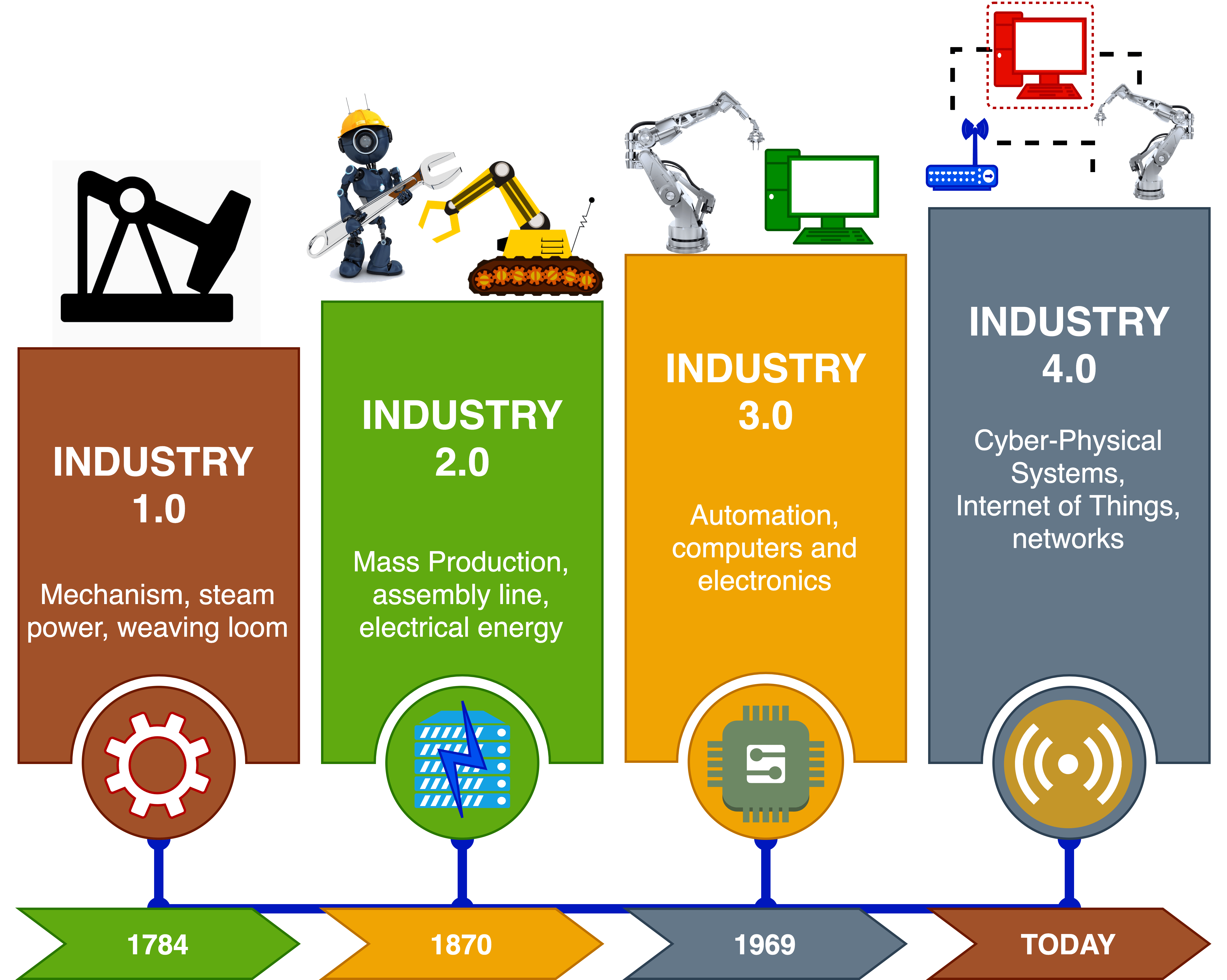}  
\caption{Revolution of smart industries.\label{revolution}}  
\end{center} 
\end{figure}

The recent technological advancements in hardware, software, and wireless communication have facilitated the emergence of a new concept termed as the Internet of Things (IoT), simplifying the human lifestyle by saving time, energy, and money \cite{gilchrist2016industry,sisinni2018industrial,greengard2015internet,Saeed2019U, gubbi2013internet, al2015internet}. In IoT, the term ``Things" represents smart devices, such as sensors, machines, and vehicles, with intelligent processing capabilities. Many interesting IoT applications include smart cities, smart homes, healthcare monitoring, intelligent transportation, smart power generation, and agriculture \cite{boyes2018industrial,liang2020toward,zarei2016internet,atzori2010internet,branger2015automated}.  
Nevertheless, the IoT technology also plays a  vital role in modern industries by introducing automation and reducing expenses. Adding IoT sensors to the industrial systems helps operators to monitor equipment in near real-time with high reliability. The Industrial IoT (IIoT) market (without the consumer IoT) is growing fast as digital transformation in many industries is accelerating \cite{sadeghi2015security,perera2015emerging,wan2016software,da2014internet,ray2018survey,li2015internet,lee2015internet}. With strong alliances between leading IIoT stakeholders and proven IIoT applications, inspire companies worldwide to invest in the IIoT market, which is expected to grow up to \$123.89 billion by 2021 (see Fig. \ref{sizemarket}). 
Throughout the literature, various terms are used to describe IoT for the industry, such IIoT, smart manufacturing, and Industry 4.0 \cite{zheng2018smart,qiu2018blockchain,liu2019performance,jeschke2017industrial}. The step towards this smart production technology or IIoT is way different from past technologies; therefore, it is also called the Industrial Revolution. The newly IIoT technology will explicitly change the overall working conditions and day to day lifestyles of people \cite{khalil2020network,chen2018edge,perera2014survey}. The revolution in smart industry from Industry 1.0 to 4.0 is illustrated in Fig. \ref{revolution}. Generally, IIoT networks consist of interconnected intelligent industrial components to accomplish high production with low cost. This is possible by real-time monitoring, precise execution of tasks, and efficient controlling of the overall industrial procedures \cite{khan2020industrial, karmakar2019industrial,zhong2017intelligent,li2017applications}. 

As the number of internet-connected devices grows exponentially, a significant amount of data is generated. In the case of IIoT, both volume of the data and its characteristics needs proper consideration because the performance of IIoT applications is strictly related to the intelligent processing of the big data, which comes from different real-time resources \cite{mahdavinejad2018machine,wuest2016machine, arachchige2020trustworthy,  wang2016green,wang2015big,hu2015system,najafabadi2015deep,liang2020towards}. Therefore,  big data analysis in IIoT networks requires intelligent modeling that can be achieved using deep learning techniques. 

As we discussed above, the most critical factor in automation and intelligence for IIoT is data modeling, analyzing, and support real-time processing. Among data analysis schemes, Deep Learning (DL) used in regression, classification, and forecasting has shown the most potential for IIoT systems \cite{liang2020towards,ahuett2018brief,ullah2019cyber}. DL approaches have benefits to learn from the given data automatically, identify the patterns, and make some precise decisions. With the advancements provided by DL methods, IIoT will transform into highly optimized facilities \cite{yan2018industrial}. The advantages of using DL include lower operation costs, no change with variations in consumer demands, enhancing productivity, downtime reduction, obtaining a better perspective of the market, and extracting valuables from the operations \cite{wang2018deep,zhang2018adaptive}. 

With all these advantages of DL in IIoT, this paper aims to present a state-of-the-art review for DL techniques, such as Convolutional Neural Networks (CNNs), Auto Encoder (AE), Recurrent Neural Network (RNN), Restricted Boltzmann Machine (RBM) and its variants, and their usage in IIoT. Notably, the DL-enabled advanced analytics framework can realize the opportunistic need for smart manufacturing. First, we review the basic DL techniques and their applications in IIoT. Then, we introduce various use cases of IIoT, where DL can significantly improve the performance of predictive maintenance, assets tracking, smart metering, remote healthcare, power-and-smart grid industry, telecommunications, and human resource management. Finally, we provide different research challenges and future trends for DL-based IIoT.

The rest of the paper is organized as follows. Key DL concepts in the context of smart industries are illustrated in Section II. In Section III, we present numerous use cases of DL in different smart industries.  Section IV provides various challenges faced by DL-based IIoT, followed by future research directions in Section V. Finally, conclusions are drawn in Section VI.

\section {Deep Learning for IIoT}
It is crucial to use smart manufacturing, making both manufacturing and production intelligent, which can have several advantages in IIoT \cite{zeng2019boomerang,chen2017integrated}. Recently, IIoT solutions are growing significantly, especially for sensor-based data consisting of various structures, formats, and semantics \cite{yao2017intelligent,lee2019deep}. Data released from IIoT technologies is derivative from various resources across manufacturing that includes product line, equipment-and-processes, labor operation, and environmental conditions. Therefore, data modeling, labeling, and analysis play an important role in making manufacturing much smarter \cite{al2020survey,lecun2015deep}. They are essential parts of handling high-volume data and supporting real-time data processing. \par

\begin{figure}[h!]
\begin{center}  
\includegraphics[width=1\columnwidth]{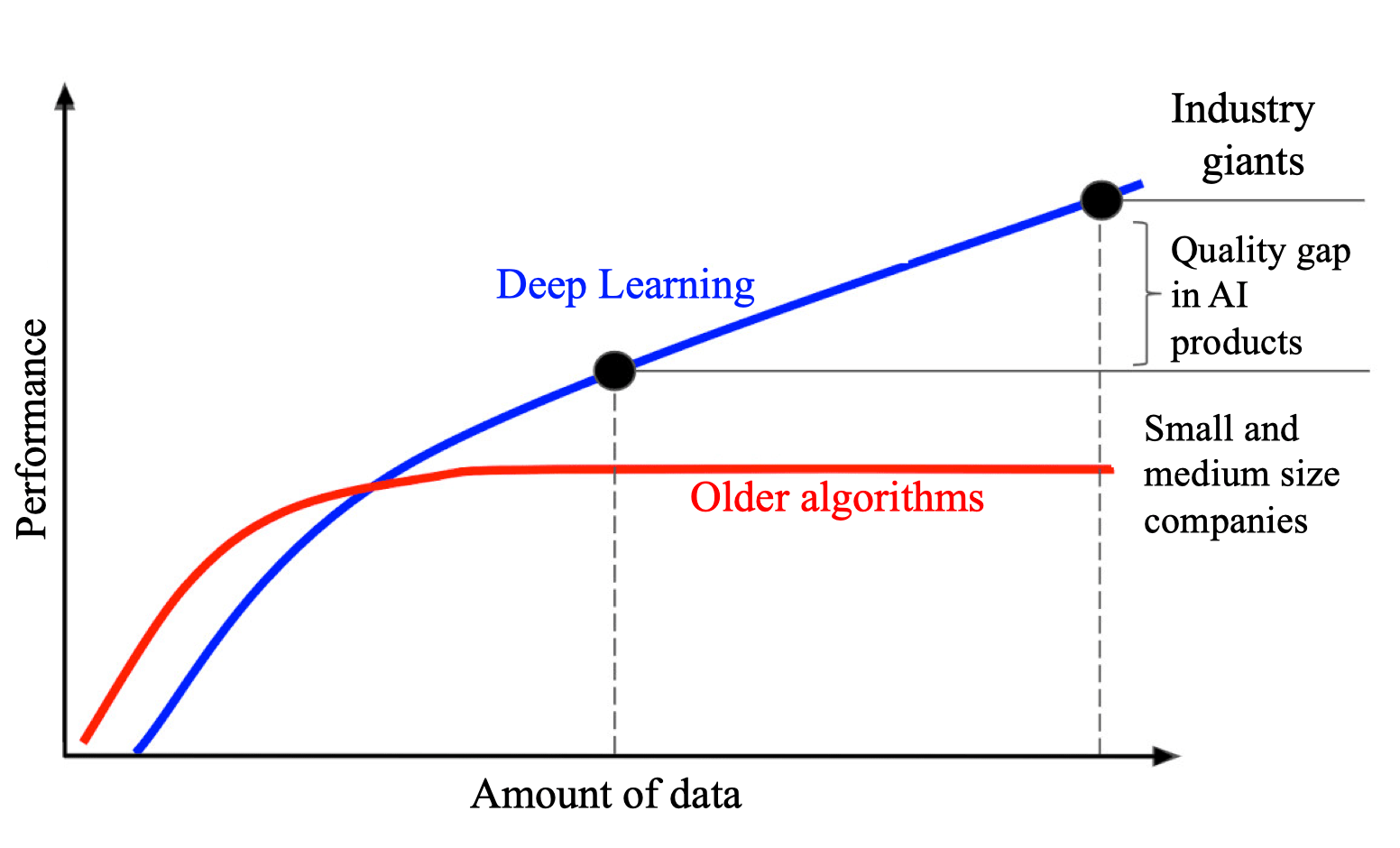}  
\caption{Comparison of DL with traditional algorithms for IIoT .\label{Fig00}}  
\end{center} 
\end{figure}

Deep Learning (DL) is one of the powerful methods of machine learning (ML) which can upgrade manufacturing into highly optimized smart facilities by processing a bunch of data with its multi-layered structure. The DL approaches can provide computing intelligence from unclear sensory data, resulting in smart manufacturing \cite{espitia2020novel}. DL techniques are useful due to their automatic data learning behavior, identifying the underlying patterns, and making smart decisions. One of the advantages of DL over traditional machine learning methods is that feature learning is done automatically, and there is no need to design a separate algorithm to do this part \cite{chen2019emerging, zantalis2019review}. The comparison of DL with traditional techniques in IIoT can be analyzed in Fig. \ref{Fig00}. 

\begin{figure*}[htp!]
\begin{center}  
\includegraphics[width=0.95\textwidth]{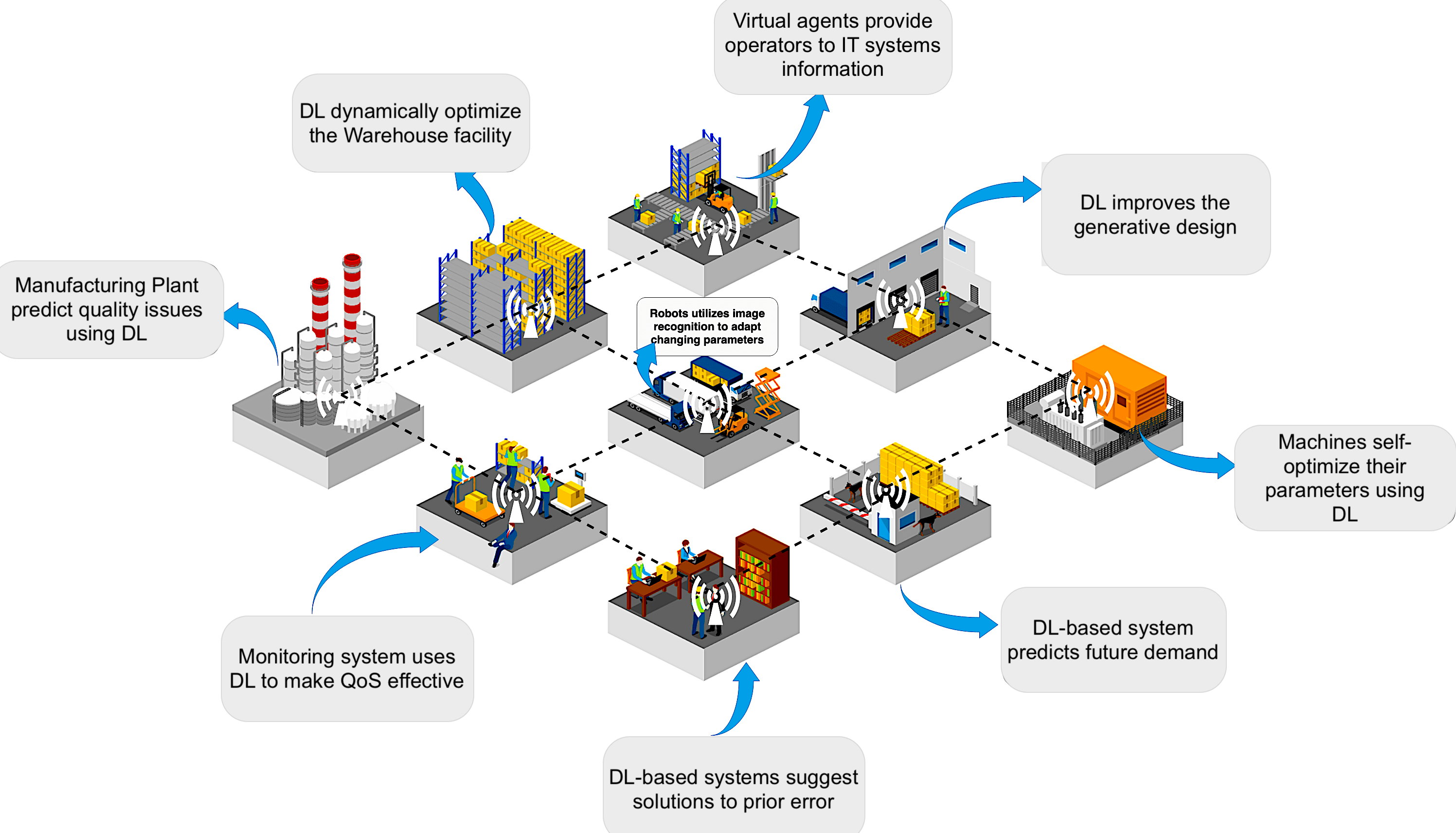}  
\caption{Deep learning in future Industrial Internet of Things.\label{Fig01}}  
\end{center} 
\end{figure*}

Different data analysis methods can be utilized to interpret the IIoT data, such as predictive analytics, descriptive analytics, diagnostic analytics, and prescriptive analytics \cite{lade2017manufacturing, Saeed2018, al2018context,liu2018blockchain,weyrich2015reference}. After capturing and analyzing the product's condition, environment, and operational parameters, we can arrive at a summary of what happens; this is called descriptive analytics. Predictive analytics utilizes statistical models and provides the prospect of making future equipment fabrication or degradation based on the provided historical data. We use diagnostic analytics to figure out the foundation cause and report the reason for failure or reducing product performance. In contrast, prescriptive analytics recommends one or more courses of action and whatever lies beyond it. Measures out of data analysis techniques are classified for developing the production outcomes or improve the problems, depicting probable result of each assessment. Fig.\ref{Fig01} shows the DL-based architecture and its impact on IIoT. In the following, we discuss various DL techniques for IIoT networks.

\subsection{Convolutional Neural Network}
\begin{figure}[htp!]
\begin{center}  
\includegraphics[width=1\columnwidth]{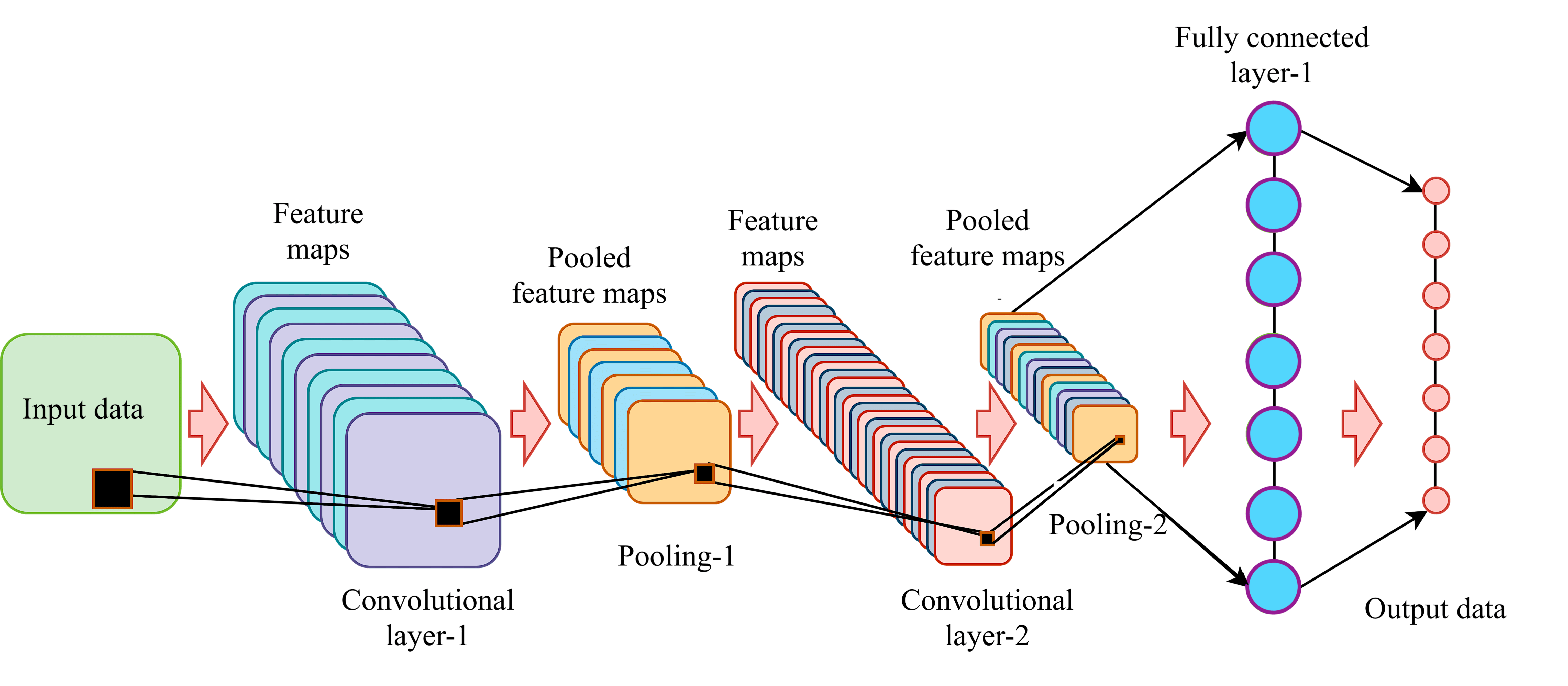}  
\caption{Generic layer-wise architecture of CNN.\label{CNN}}  
\end{center} 
\end{figure}

Convolutional neural network (CNN) consists of artificial neural network layers that was first introduced for two-dimensional natural language processing (NLP), image/signal processing and speech processing \cite{khalil2019speech,li2017deep}. CNNs are a type of DL method that convolve different neural layer for a high dimensional input data, arriving at a particular lower dimensional output. The CNNs follows a multi-layer feed-forward architecture where the whole procedure of data analysis-and-managing includes filtering and dimension reduction by using convolutional and pooling layers. Considering the generic layer-wise architecture of CNN with its convolutional, pooling and fully-connected layers are depicted in Fig.\ref{CNN}.

Moreover, recently, CNN is explored and utilized for 1D consequent data analysis, incorporating various IIoT applications \cite{weimer2015context,li2018incremental}. The feature learning in CNN is usually obtained by interchanging and assembling the convolutional layers and pooling processes. These layers (convolutional layers) convolve with the given input data in raw form utilizing local procreative invariant features and multiple local kernel filters. On the other hand, the pooling layers extricate the signified features with fixed-length sliding windows of the raw form input data. This is carried out using various pooling techniques such as average-pooling and max-pooling \cite{wen2018fault,he2017deep,subakti2018indoor}. The average-pooling basically measures the mean value of the specified section and considered it as the pooling value of that section. In contrast, the max-pooling chooses the maximum value for a stipulated section of the featured map being the utmost substantial feature. It should be noted that average-pooling is not optimal in all feature extraction scenarios, however, the max-pooling is reasonably well-matched for sparse feature extraction.

After successful multi feature-learning, the fully connected layers transform the 2D feature mapping into a 1D vector and feed it into activation functional layer for better model representation. In CNN, the most commonly used activation functional layer is softmax, which usually transforms the output of the previous layer into the most suitable and essential output probabilistic distribution. This layer-wise architecture is quite helpful in IIoT applications such as the detection of surface defects that occurs in the manufacturing process \cite{neogi2014review, monsone2019CNN,wang2019machine}. The CNN is mainly trained by using Gradient-based backpropagation that minimizes the overall minimum-mean-squared-error (MMSE) and cross-entropy (CE) function for loss occurrence. The CNN has numerous advantages as compared to its counter parts' neural networks, including parameter distribution with fewer numbers, local connectivity with sparse interfaces, and equid-variant illustration, which is mostly invariant to object localities. 

The CNN is advantageous in IIoT, as it provides extensive knowledge (in terms of feature extraction) using various datasets with minimal human supervision \cite{liang2020towards}. This not only enables the industrial process to be precise but can also effectively identify the underlying defects using its visual defect detection capability \cite{lee2020integration,posada2015visual}. Furthermore, the requirement of hand-crafted features, lengthy trials of various procedures, and error occurrence are reduced due to the accurate feature measuring capability of CNNs towards industrial processes.

\subsection{Auto-Encoders}
\begin{figure}[h!]
\begin{center}  
\includegraphics[width=0.775\columnwidth]{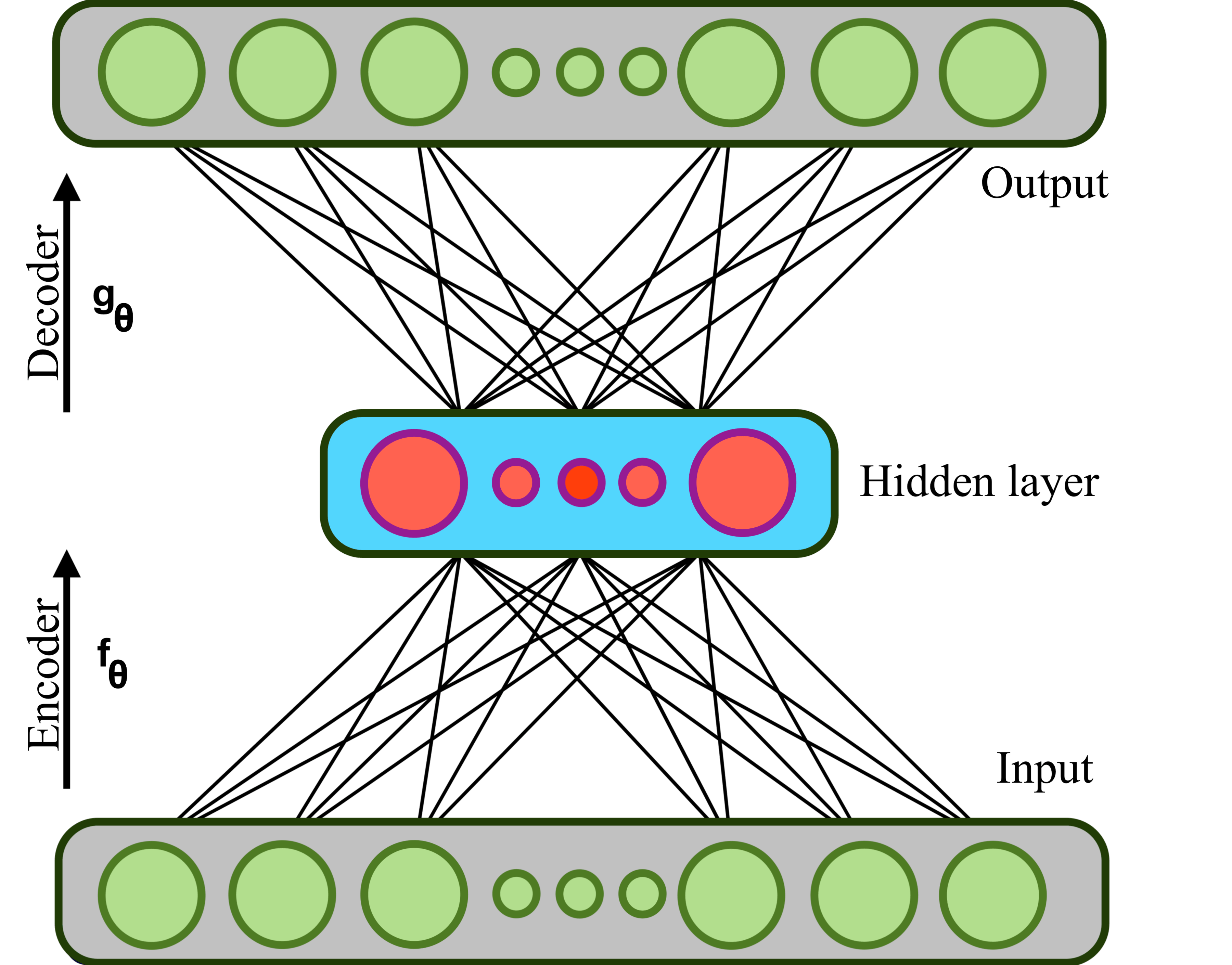}  
\caption{Generic layer-wise architecture of AE.\label{Fig4.AE}}  
\end{center} 
\end{figure}

Auto Encoder (AE) is an unsupervised learning neural network which extricates various features from the provided input data without any labeled information requirement \cite{wong2018recurrent}. The AE is primarily comprised of an encoder $f_{\theta}$, a decoder $g_{\theta}$, and a hidden layer in between, as shown in Fig. \ref{Fig4.AE}. The encoder $f_{\theta}$ is capable of performing data compression by mapping the given input to a defined hidden layer in cases where high dimensionality is desired, and the decoder can reconstruct the approximation of the input information \cite{muna2018identification}. Further, it can effectively reconstruct the approximate relative to the input information. The AE can perform comparably to principal component analysis (PCA) if the activation function is linear with few defined hidden layers. It is observed that if the input data is non-linear, then more hidden layers need to be defined to create a deeper AE \cite{ren2018bearing}. Usually, Stochastic gradient descent (SGD) is utilized to measures various factors and construct an AE with minimal loss in the objective function. This loss is minimized in terms of cross-entropy loss or least square loss \cite{essien2020deep}.
The most important feature that has made the AE useful for the IIoT application is its ability to reduce the input data dimensions and learn the features of non-label data \cite{wen2017new}. AE can also help in the security of IIoT networks as it is capable of providing information regarding any intrusion detection in the IIoT environment \cite{huang2019towards,hassanzadeh2015towards,pescatore2014securing}. Hence some precautionary measures can be taken to avoid any mishap in the industrial process such as safety, production, and warehousing. 

\subsection{Recurrent Neural Network}
\begin{figure}[h!]
\begin{center}  
\includegraphics[width=1\columnwidth]{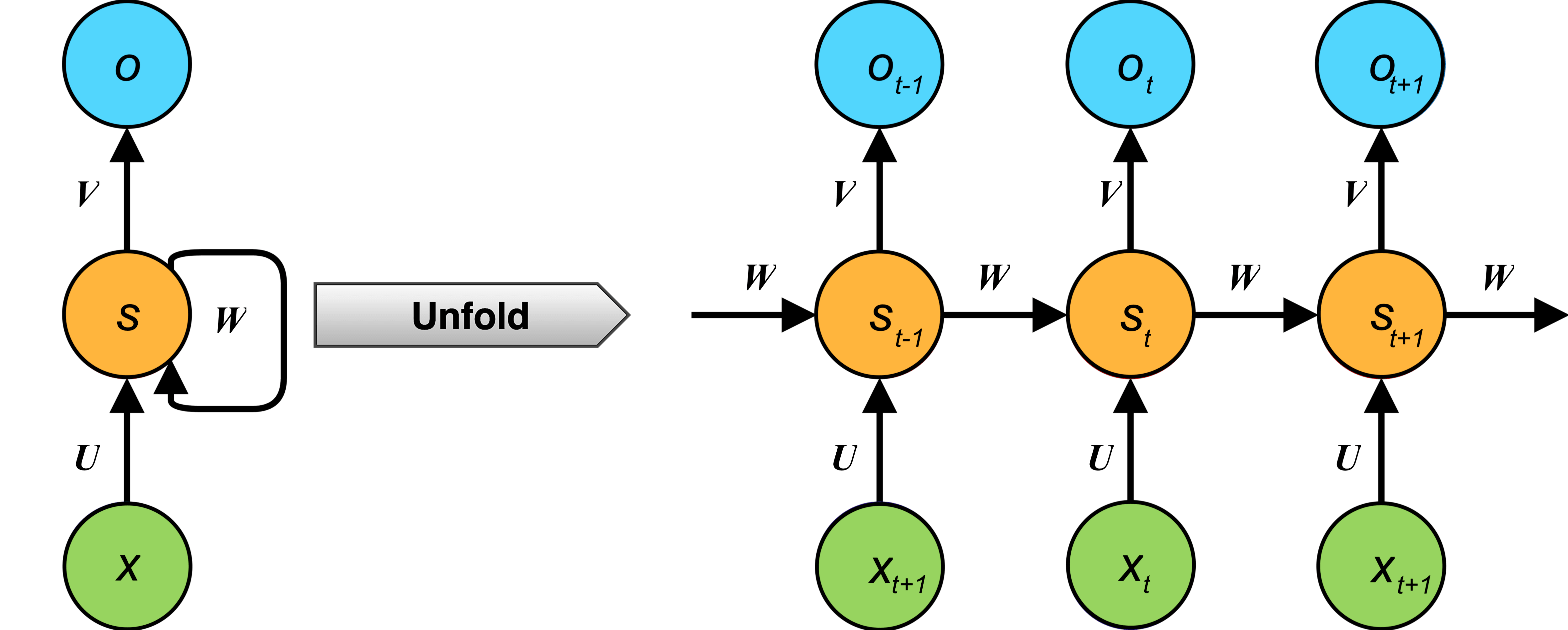}  
\caption{Generic layer-wise architecture of RNN.\label{Fig5RNN}}  
\end{center} 
\end{figure}

Recurrent Neural Network (RNN) has an exclusive capability of hierarchical links among the neurons for modeling sequential data \cite{sherstinsky2020fundamentals,putchala2017deep}. Fig. \ref{Fig5RNN} provides the basic architecture of RNN with inputs, outputs, and underlying deep hidden layers. The input variables are expressed as $x_{t-1}$, $x_{t}$ \text{and} $x_{t+1}$ (depending upon the number of input variables), $s_{t}$ provides the information of underlying hidden states, whereas  $o_{t-1}$, $o_{t}$ \text{and} $o_{t+1}$ gives the respective outputs at time instant $t$. The terms $\boldsymbol{U}$, $\boldsymbol{V}$, and $\boldsymbol{W}$ are the respective hidden matrices whose values vary for each timestamp where , the hidden states can be measured as $\boldsymbol{S}_{t}=f(\boldsymbol{U}_{x_{(t)}}+\boldsymbol{W}_{S_{(t-1)}})$ \cite{khalil2019speech}. Thus, RNNs are appropriate to acquire precise measurements from the sequential data, allowing it to persist the information in hidden layers and capture the preceding conditions of the data. Nowadays, a reorganized form of RNN is used in a layer-wise architecture that can effectively calculate the variance between various time steps \cite{lepenioti2020machine,roy2018deep}. The RNNs take the data as a sequential input in the form of a vector where the existing hidden state is computed by using an activation function, such as tan$h$ or sigmoid. The initial portion of the data is calculated with the provided input; however, the next section is acquired form the underlying hidden state at the earlier step time $t$ \cite{temeng1995model}. The targeted output $o_t$ is then measured with the help of these hidden layer states through softmax or any other available method. Once the whole order of data is processed, the hidden states represent the learning illustration of the whole consecutive input data. Further, a traditional multilayer-perception (MLP) is added on the topmost layer to map the target representations. 

Unlike traditional neural networks, RNNs use back-propagation-through-time (BPTT) for model training. The RNN is usually unrolled in time at the initial stage, and later on, which is then used as an extra layer to the whole architecture. Subsequently, the back-propagation is used to measure the gradients decent \cite{wu2018approach,chang2018review}; however, the BPTT undergoes the shattering problem when applied during model training. The reason for this is that RNN is not durable enough and has complications to capture the long-term dependencies effectively. In literature, a diverse set of improvements are proposed to tackle these issues, amongst which is the long short-term memory (LSTM) solution \cite{lu2008adaptive}. The fundamental knowledge of LSTM is the cell state, which is responsible for allowing the data to be fluttered down with a linear interface. In comparison with single state RNN, in LSTM, different gates such as input, forget, and output gates are utilized to regulate the cell state. This facilitates each recurrent unit to apprehend the long-term dependences for every time scales adaptively.

Initially, RNNs were applied effectively in NLP and textual analytics, using their temporal behavior to rely on time series. However, the same knowledge can be applied to various industrial applications such as optimization of smart manufacturing, where the computational load can be reduced with the help of RNN \cite{pacella2007using}. This will improve the manufacturing process without any sacrifice of predictive power usage. In comparison to NLP, the RNN can effectively be used in IIoT to predict potential issues that arise with different machine health parameters. It can be helpful in smart factories where future predictions can minimize the overall manufacturing cost and enhance downtime of the product.

\subsection{Restricted Boltzmann Machine}
\begin{figure}[h!]
\begin{center}  
\includegraphics[width=0.49\columnwidth]{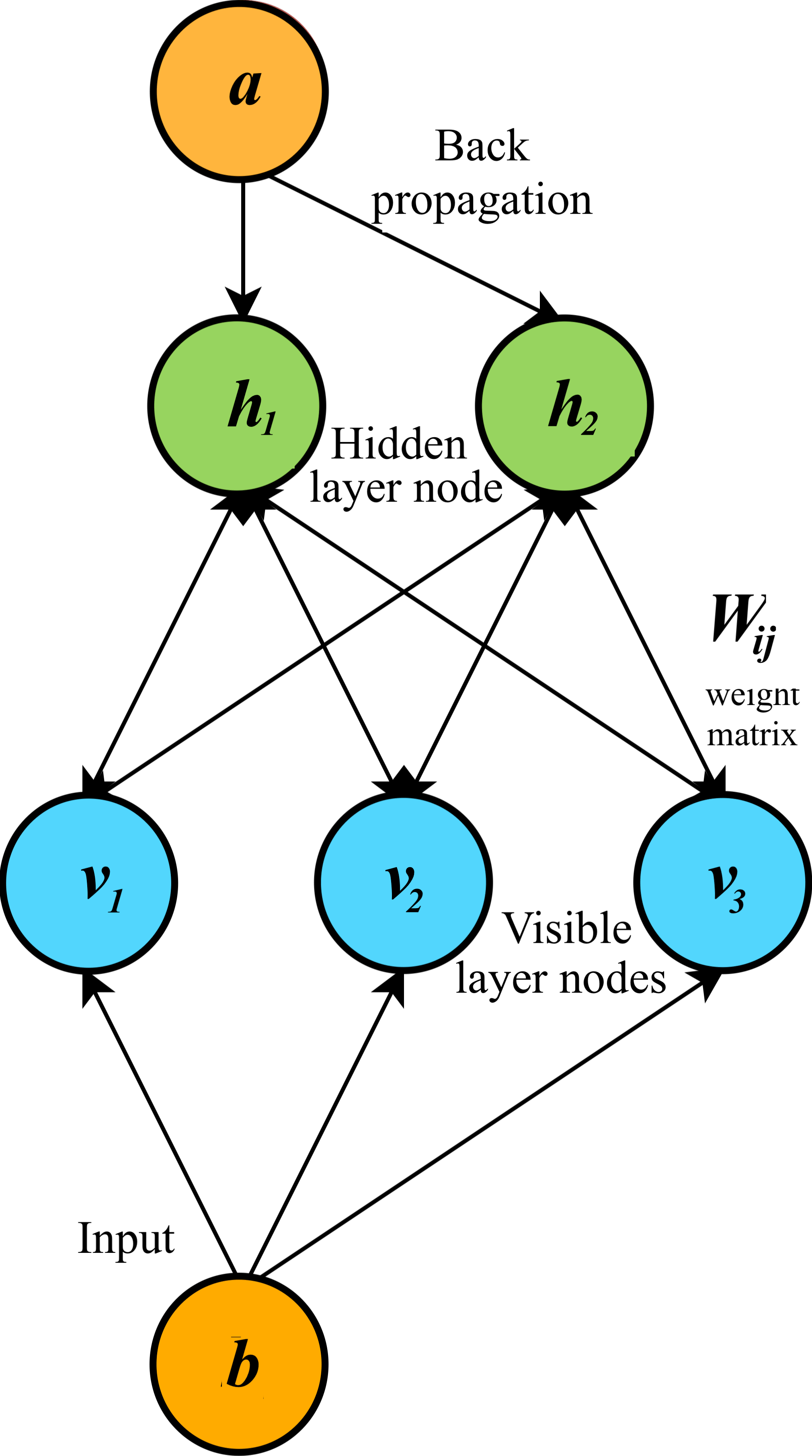}  
\caption{Generic layer-wise architecture of RBM.\label{Fig12.RBM}}  
\end{center} 
\end{figure}

Restricted Boltzmann Machine (RBM) is an artificial neural network mainly with two-layer networking architecture that includes a hidden layer and a visible layer, respectively. In RBM, only the hidden and visible layers are connected; however, there is no correlation between the neurons of the identical layer \cite{zhang2018overview}. It is a kind of energy-based technique, where the visible layer is responsible for input data, while the hidden layer is utilized for different feature extraction. Additionally, it is assumed that all the hidden nodes are conditionally self-determining and have no interdependency \cite{bai2018manufacturing}. The offsets and weights of the visible and hidden layer are revolved throughout iterations to sort the visible layer an estimate to that of input data \cite{saleem2019deep}. Conclusively, the hidden layers deemed as an alternate depiction of the corresponding visible layer. A generalized layer-wise architecture of RBM is given in Fig. \ref{Fig12.RBM}.
The factors in hidden layers are utilized to characterize the input data in the realization of dimensionality reduction and various data coding mechanisms. Supervised learning methods comprises naive Bayes, support vector machine (SVM), linear regression, and belief propagation (BP) can be used for data regression and categorization. RBM usually retrieve the essential features from the training datasets adaptively and take advantage of avoiding the local minima problem. According to \cite{ma2019survey}, RBM is vastly used as a principal learning mechanism on whose basis other variants are modeled.

\subsubsection{Deep Belief Network} 
\begin{figure}[h!]
\begin{center}  
\includegraphics[width=1\columnwidth]{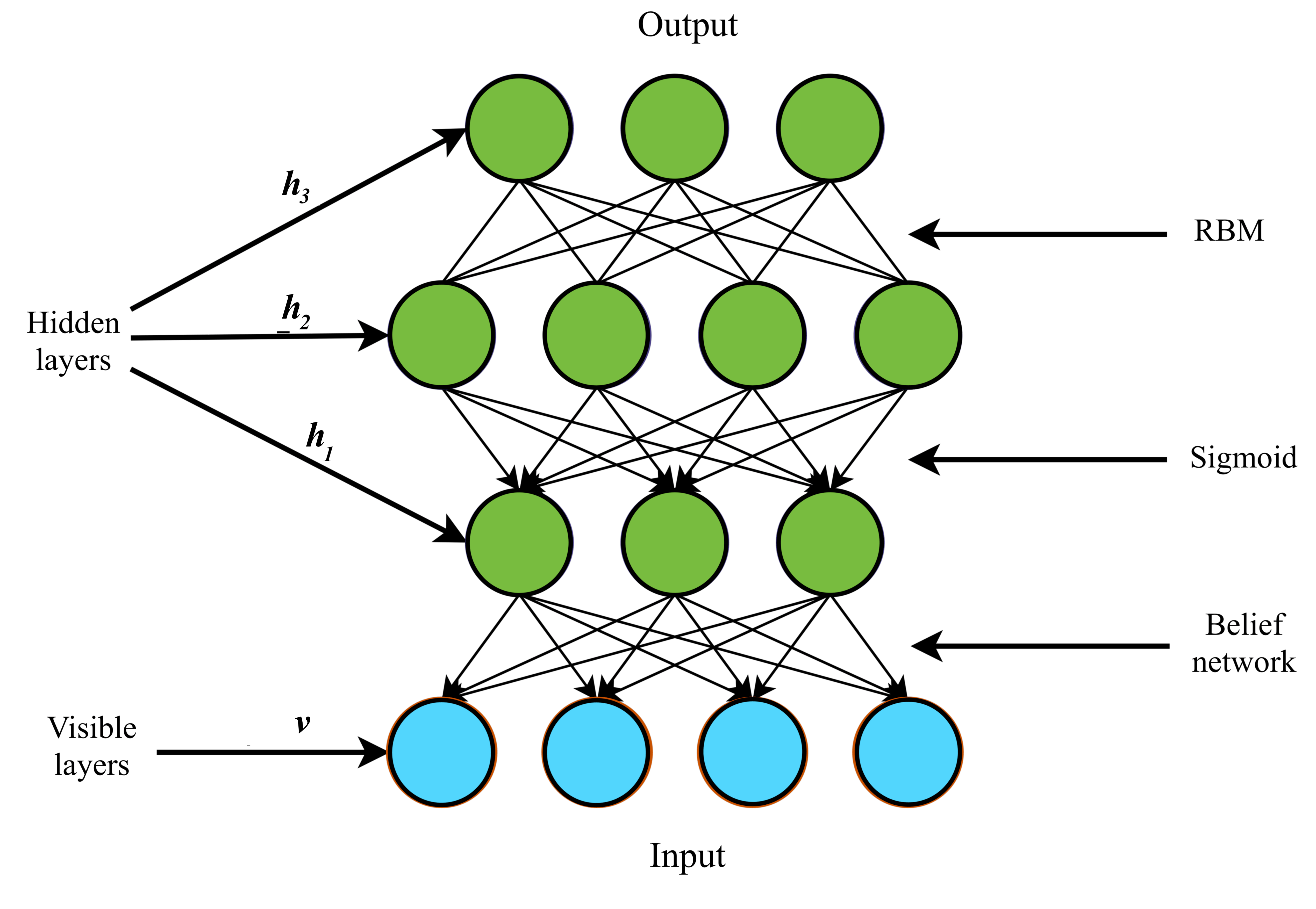}
\caption{Generalized layer-wise architecture of DBN.\label{Fig7.DBN}}  
\end{center} 
\end{figure}

Deep Belief Network (DBN) is a variant of RBM and is fabricated by mounding various RBMs. In DBN, there are input layer units, hidden layer units, and the output layer units. The underlying architecture of DBN is depicted in Fig. \ref{Fig7.DBN}. A fast-greedy algorithm is used to train the DBN, and a wake-sleep algorithm is utilized to fin-tuned the different parameters in its deep architecture \cite{banjanovic2020intelligent,alrawashdeh2016toward}. The area closed to the visible layer is tackled by Bayesian Belief Network (BBN), and far away region is investigated by the RBMs \cite{ranzato2008sparse}. It can be observed that the lowermost layers in DBN are directed while the uppermost layers are undirected. 

According to \cite{xu2018identification}, the DBNs are quite beneficial in pre-training due to their unsupervised nature, especially for unlabeled and massive datasets. Moreover, the DBNs is also advantageous due to their low computational capability. Nevertheless, the DBNs approximation method follows bottom-up fashion, where the greedy layer is responsible for learning only single layer features and is unable to fine-tune with the residual layers \cite{xu2015temporally,huda2018malicious}.

\subsubsection{Deep Boltzmann Machine} 
\begin{figure}[h!]
\begin{center}  
\includegraphics[width=0.835\columnwidth]{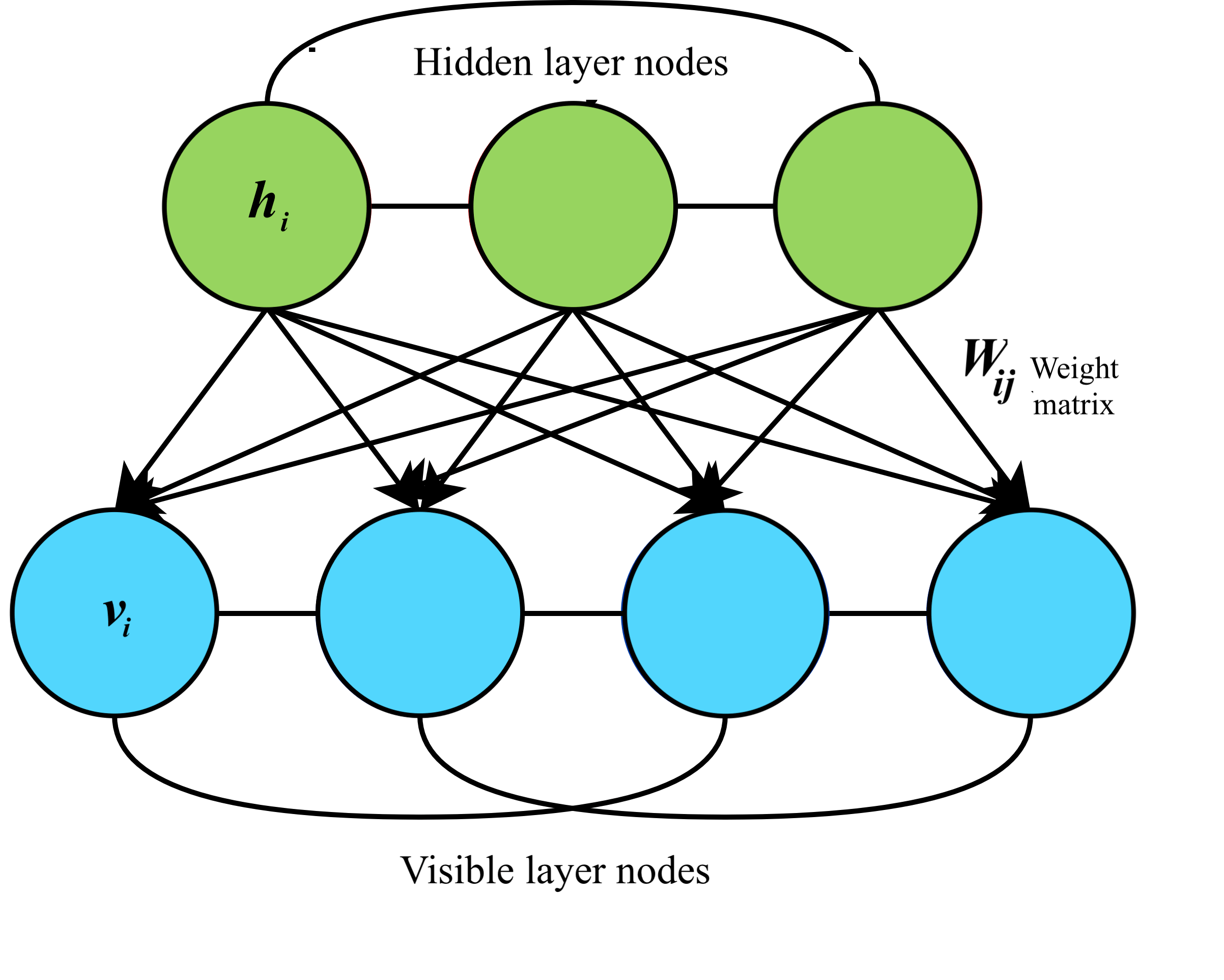}  
\caption{Generalized layer-wise architecture of DBM.\label{Fig6.DBM}}  
\end{center} 
\end{figure}

Deep Boltzmann Machine (DBM) is another variant of RBM and is also known as deep-layer structured RBMs. In DBM, the hidden layer units are clustered into a deeper hierarchical layer structure. There is full-fledged connectivity between the adjacent two layers; however, there is no connectivity between the nodes within a specific layer or amongst the neighboring layers, as shown in \ref{Fig6.DBM}. Due to the assembling of multi RBMs in DBM, it is capable of ascertaining complicated edifices and paradigm high-level illustration of the provided input data \cite{salakhutdinov2009deep}. DBM is a complete undirected model as compared to DBN, which is both undirected/directed in nature. Also, the DBM model is typically trained simultaneously and is having a high computation cost. In contrast, the DBN can be adequately trained in a layer-wise manner and is computationally less expensive.

The RBM techniques can be quite useful for IIoT by improving their efficiency, accuracy in production, safety in operations, and reliability in the overall system. For instance, recently in \cite{wang2019deep}, a hybrid model on the basis of Gaussian-Bernoulli deep Boltzmann machine (GDBM) was used for optimal detection processes such as manufacturing or production in a smart industrial setup. The RBM and its variants can be utilized in IIoT for other tasks that include material handling, optimized designing, an inspection of the product quality, and forecasting.

\subsection{Recursive Neural Network}
\begin{figure}[h!]
\begin{center}  
\includegraphics[width=0.92\columnwidth]{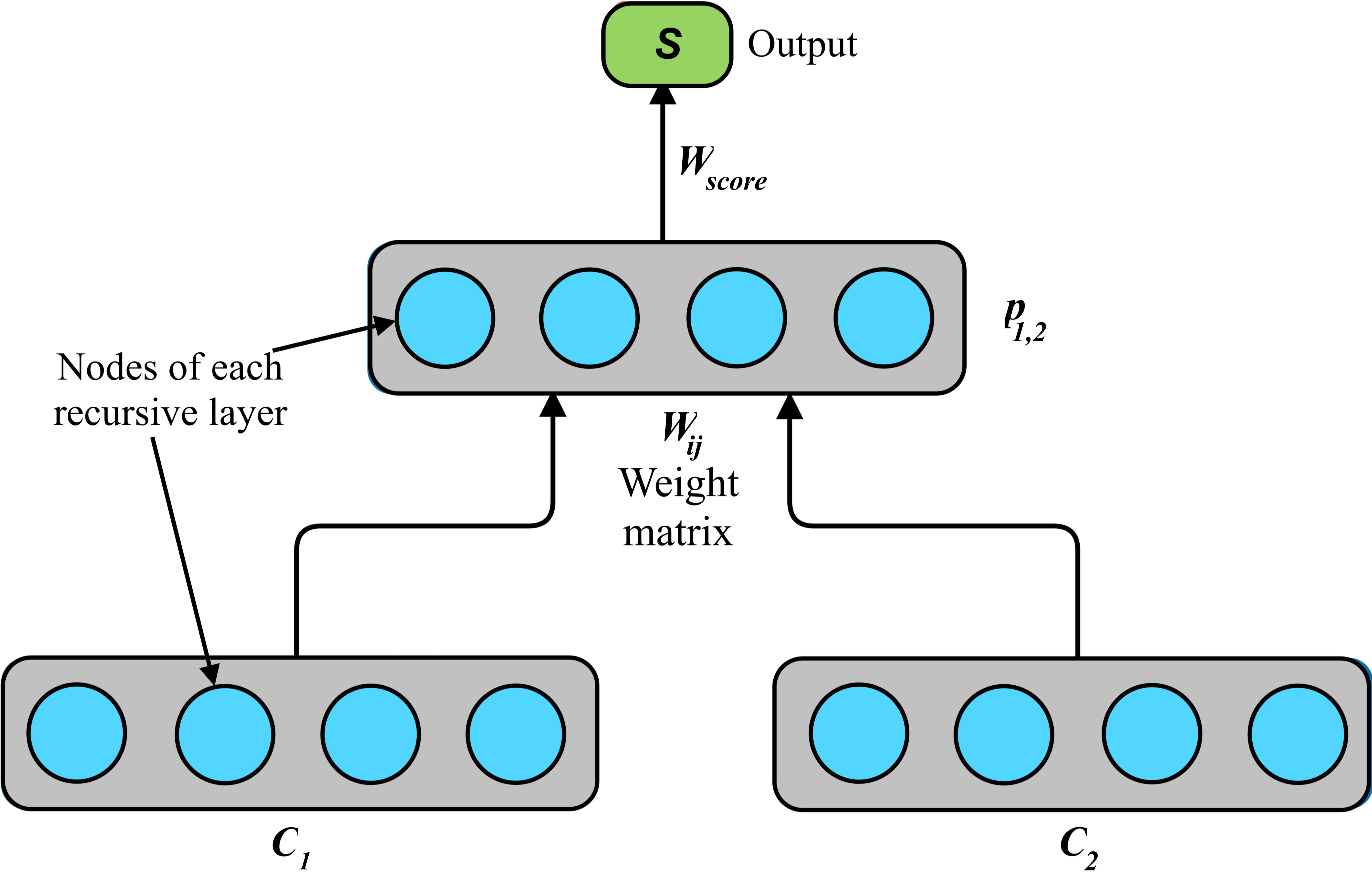}  
\caption{Layer-wise architecture of RvNN.\label{Fig8.RvNN}}  
\end{center} 
\end{figure}

Recursive Neural Network (RvNN) is also a well-known DL method that does not require any tree-structure sequence as an input \cite{socher2011parsing, dong2014adaptive, wu2017long}. It can understand the parse-tree sequences for the given input data and categorize it. RvNN is created with a set of similar weights applied recursively to the input data \cite{irsoy2014deep}. The weights are defined almost for each node and are not restricted to a specific node. Furthermore, the RvNN is categorized as a classical architecture used to operate on inputs in a structured form, predominantly, on guided acyclic graphs. A simplified layer-wise architecture of RvNN is illustrated in Fig. \ref{Fig8.RvNN},
where $p_{1,2}=\tanh(\boldsymbol{W}[\boldsymbol{C}_{2}:\boldsymbol{C}_{2}])$ is the parent matrix, the terms $\boldsymbol{C}_{1}$ and $\boldsymbol{C}_{2}$ are the respective n-dimensional vectors defined for nodes, and $\boldsymbol{W}$ is the weighted matrix of order $n \times 2n$. RvNN determines the overall score for any potential pair that can be fused to design a syntactic tree structure and then merged with compositional vectors \cite{mishra2006drought}. After these possible pairs get merged, the RvNN creates various components that include the bounded territory characterizing vectors and the classification labels. 

RvNN-based DL methods can efficiently perform the tasks of natural language processing, image processing, and speech processing \cite{kamp1990recursive, chinea2009understanding}. 
Nevertheless, it can also effectively handle IIoT applications, including smart manufacturing, smart object detection during assembling, packaging, and machine controls \cite{han2019industrial, he2018statistical, nowopolski2017recursive}. It can also be used in controlling industrial pollution by providing an earlier forecast with greater precision \cite{biancofiore2017recursive}.

\subsection{Comparison of DL models}
From the above discussions, it is clear that CNN and RNN are complex algorithms for learning representations and modeling. On the other hand, RBM and AE are used in a layer-by-layer fashion for the pre-training of a neural network to illustrate the given input data. The top layer in all of these DL techniques represents the targets. The Softmax layer is used when these targets are discrete; however, linear regression is utilized for continuous targets. Some of these DL techniques are dependent on data labeling that includes RBM, AE, and their variants, which are called unsupervised or semi-supervised learning. In contrast, the other methods, including CNN, RNN, and RvNN, are supervised due to their non-dependency on labeling input data. The advantages and limitations of these DNN techniques, along with their applications, are presented in Table \ref{tab:table1}.
\newcolumntype{C}{>{\arraybackslash}X} % centered version of "X" type
\setlength{\extrarowheight}{0pt}
\begin{table*} [htp!]
 \caption{Comparison of Deep learning models in context of IIoT}
\label{tab:table1}
\begin{tabularx}{\textwidth}{@{}l*{10}{C}c@{}}
\toprule
$\textbf{DL model}$ & $\textbf{Applications}$ & $\textbf{Advantages}$ & $\textbf{Limitations}$ & $\textbf{References}$   \\ 
\midrule
CNN & Feature learning by stacked convolutional and pooling layers & minimizes the shifting, scalability and alteration & higher hierarchical models requires complex computations & \cite{khalil2019speech,li2017deep,weimer2015context,li2018incremental,wen2018fault,he2017deep,subakti2018indoor,neogi2014review, monsone2019CNN,wang2019machine,liang2020towards,lee2020integration,posada2015visual} \\
AE & Encoding do the unsupervised learning and dimensionality reduction in data & preserve only meaningful input data while the irrelevant data is filtered & Sparse illustration and layer-by-layer error procreation are not assured & \cite{wong2018recurrent,muna2018identification,ren2018bearing,essien2020deep,wen2017new,huang2019towards,hassanzadeh2015towards,pescatore2014securing} \\
RNN & Temporal outline in recurrent links while distributed states are in time-series data & temporal correlation is apprehended in progressive data with short-term info retention & Model training is difficult to keep long-term dependencies & \cite{sherstinsky2020fundamentals,putchala2017deep,khalil2019speech,lepenioti2020machine,roy2018deep,temeng1995model,wu2018approach,chang2018review,lu2008adaptive,pacella2007using}  \\
RBM & Connectivity between input and output is elaborated by hidden layer variables & robust to input ambiguity, and pre-training stage do not require label training & take much time to execute optimization of parameters & \cite{zhang2018overview,bai2018manufacturing,saleem2019deep,ma2019survey,banjanovic2020intelligent,alrawashdeh2016toward,ranzato2008sparse,xu2018identification,xu2015temporally,huda2018malicious,salakhutdinov2009deep,wang2019deep}  \\
RvNN & Able to understand the parse-tree sequences in the provided data & Has the capability to capture long-distance dependencies & The parsing is slower and usually have domain dependency & 
\cite{socher2011parsing,dong2014adaptive,wu2017long,irsoy2014deep,mishra2006drought,kamp1990recursive, chinea2009understanding,han2019industrial, he2018statistical,nowopolski2017recursive, biancofiore2017recursive} \\
\bottomrule
\end{tabularx}
\end{table*}
\vspace{1.5em}

\section{Key Use Cases of DL-based IIoT}
In this section, we describe potential opportunities for DL-based IIoT, which can handle the communication and resources of underlying smart nodes. One of the necessary parts of modern smart manufacturing is computational intelligence, which enables accurate insights from data for efficient and better future decision-making. As we have mentioned earlier, DL is one of the top fields in the investigation of different manufacturing lifecycle stages covering concepts, design \cite{zhang2017comprehensive}, evaluation, production, operation, and sustainment \cite{lee2013recent}. These various data mining applications in manufacturing are discussed and reviewed in \cite{harding2006data}, which covers different production processes like operation, maintenance, fault detection, effective decision making, and quality improvement of the product. Similarly, the authors in \cite{esmaeilian2016evolution, kang2016smart} review the evolution of future manufacturing, where they emphasize the importance of data modeling and manufacturing intelligence analysis. In the following, we discuss various use cases of DL in IIoT networks.

\subsection{DL for Predictive Maintenance}
\begin{figure*}[h!]
\begin{center}  
\includegraphics[width=0.875\textwidth]{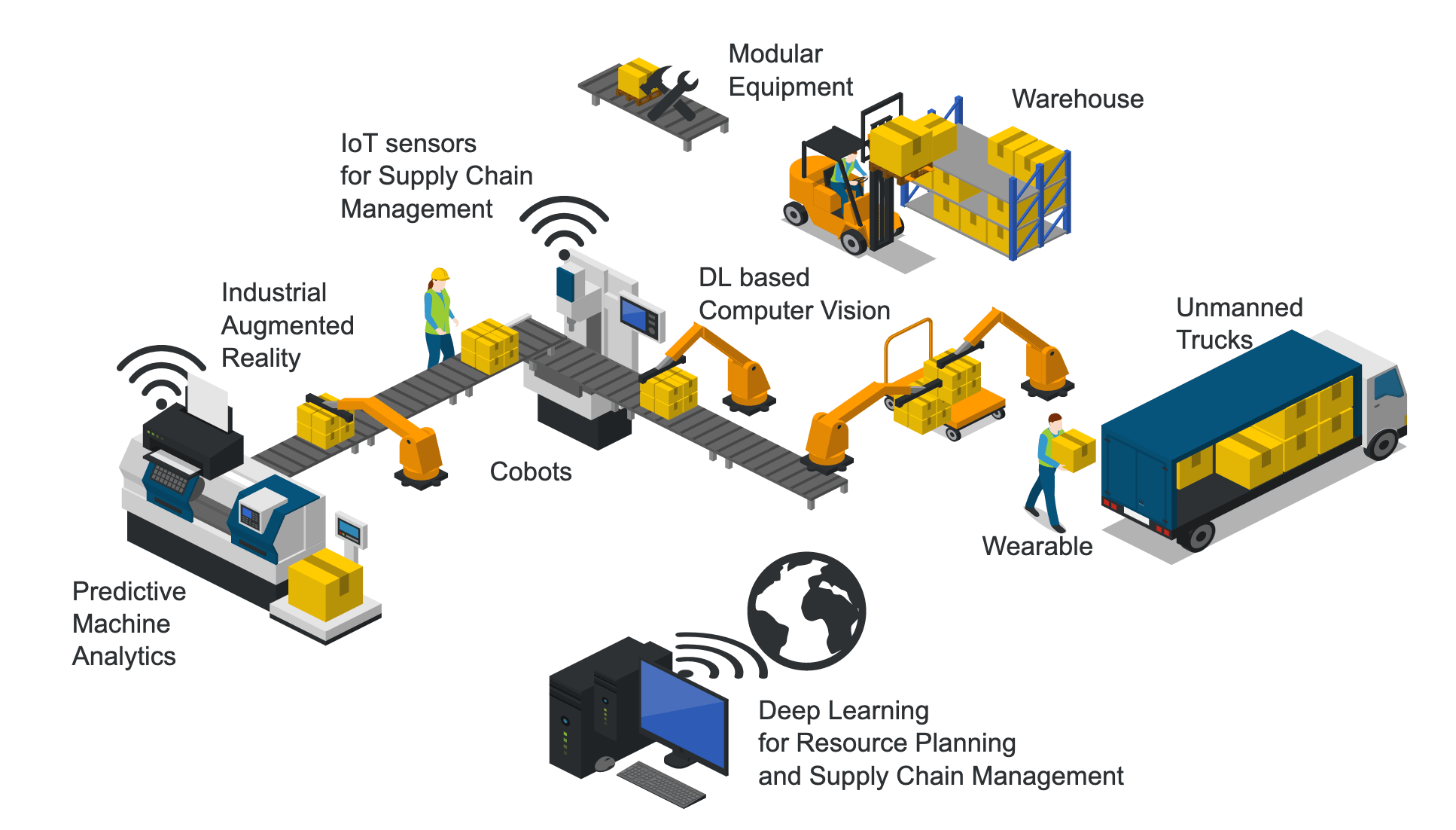}  
\caption{DL-based industrial infrastructure for predictive maintenance.\label{Fig11}}  
\end{center} 
\end{figure*}

Maintenance problems can be quite challenging and diverse in nature. Therefore, the predictive data fed to the Predictive Maintenance (PdM) unit has to be tailored for a particular problem \cite{krishnamurthy2005design,lenz2013data}. Hence, the literature on various approaches to input information for the PdM is quite rich \cite{luo2013online}, where DL techniques seem to be among the most popular \cite{su2006intelligent}. Fig. \ref{Fig11} provides the effectiveness of DL-based industrial infrastructure for PdM.

DL-based PdM solutions can be categorized as follows:
\begin{enumerate}[(a)]
\item Supervised, where the modeling dataset has the information of failures occurrence.
\item Unsupervised, where the modeling dataset has only logistics and processes information without having any maintenance data.
\end{enumerate}

The accessibility to maintenance data depends mostly on maintenance management policies. For example, in case of \textit{Run-to-Failure(R2F)} policies \cite{susto2012predictive}, the interventions of maintenance are only performed when a failure occurs. The information for a maintenance cycle, i.e., the activity between two successive failures, exists, and therefore supervised learning methods can be used. On the contrary, in the presence of PvM policies, the maintenance is performed before any potential failure, and therefore full maintenance cycle may not exist \cite{mourtzis2016industrial,civerchia2017industrial,susto2014machine}. In such a case, only unsupervised learning methods are reasonable. In general, supervised learning methods are preferable in industries because of the extensive use of R2F policies.

Additionally, in PdM, regression techniques are utilized when predicting the lifetime of an industrial process or equipment either directly \cite{butler2010particle} or indirectly  \cite{heng2009intelligent}. In contrast, classification-based techniques are used for PdM when discriminating between the health conditions of industrial equipment \cite{baly2012wafer}. Moreover, classification techniques can also distinguish defective and flawless processes on the basis of observed data.

\subsection{DL for Assets Tracking}
The data analysis feature and deciding on historical data of AI are being in the exploration phase by assets and wealth management firms \cite{kala2020assets,luthi2019distributed,verma2019digital}. According to the AI Opportunity Landscape, approximately 13.5\% of AI vendors in banking offer assets and wealth management solutions. Digital asset management (investment portfolio, etc.) or distributed industrial assets (like transportation machinery, i.e., truck, factory machinery) are the applications DL in assets management. It takes the data to DL-techniques and makes automatic decision-making, whether what to be best for the future.

Most of the time, the data related to financial, news, and industrial sensors are unlabeled, and unsupervised learning works best in making investment decisions on exciting new ways.  The report published in the Financial Stability Board in 2017 states that AI and ML firms were able to increase their assets over \$10 billion till 2017, and this number is expected to grow rapidly in the next five years  \cite{rohit2018iot,turner2020utilizing,seneviratne2018smart}. The main applications of asset management through DL are digital assets management like investment portfolio and advisory consumers and investment management like physical asset management, including industrial predictive asset management.

\subsection{DL for Smart Metering}
Smart meters are essentially called smart because they are intelligent and enable two-way communication with the distribution service operator (DSO) and smart appliances  \cite{marvin1999pathways,shi2017deep,wang2018deep12}.
Smart meters can measure both consumed energy the amount of energy that is injected into the network. The data is sent at regular intervals to distribution network operators who, among others, use it to map consumption peaks and gain insights, while passing it on to the energy supplier for invoicing purposes \cite{yang2019smart}. One of the smart meters' promises is a more precise measurement and invoicing method in combination with dynamic pricing. With ongoing electrification, digitalization, decentralization, and rising demand for power, along with the addition of new energy sources, smart meters are deemed critical for the electricity sector \cite{livgard2010smart,khatri2019short}.
\begin{figure*}[h!]
\begin{center}  
\includegraphics[width=0.9\textwidth]{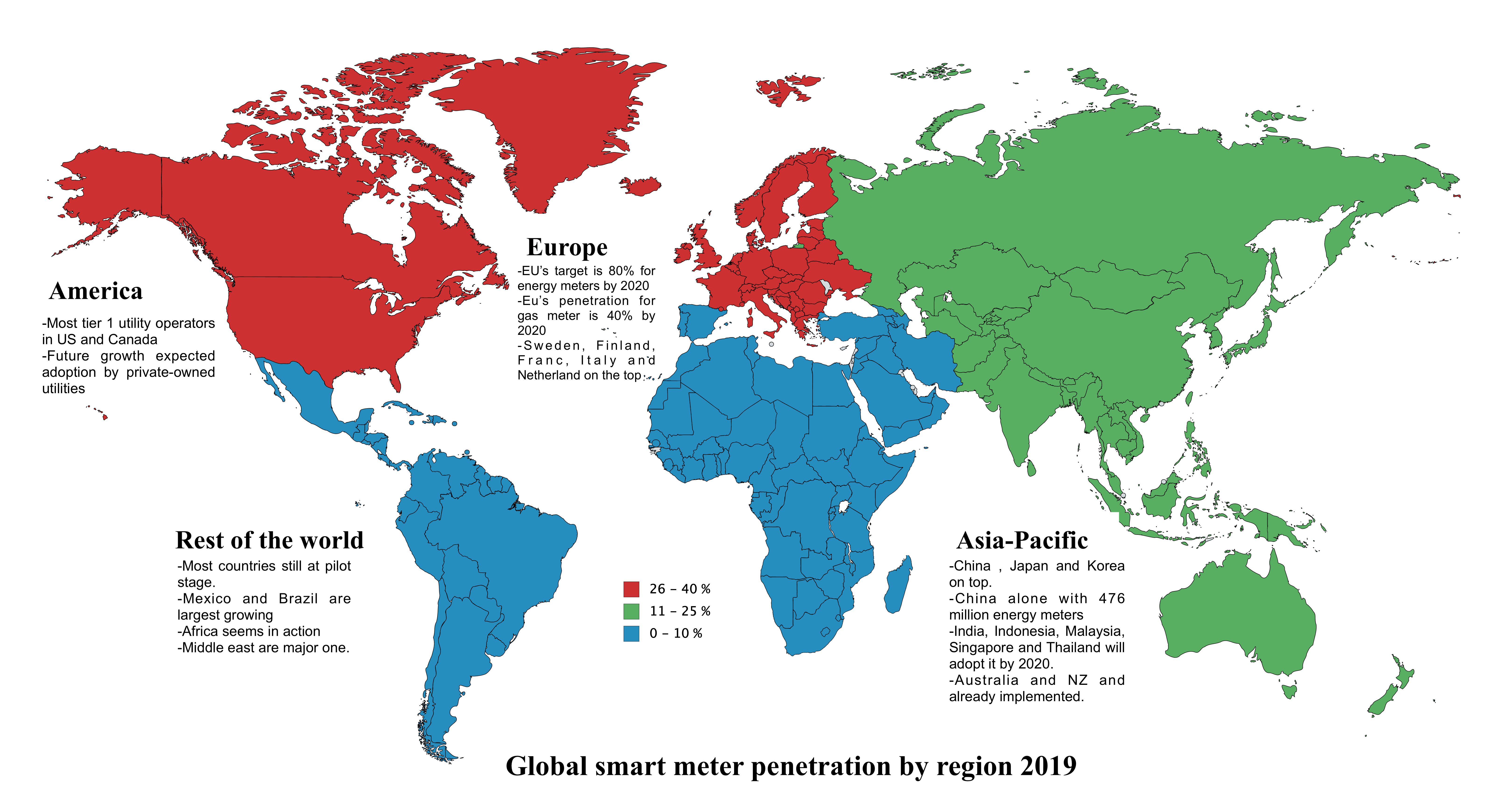}  
\caption{Smart Meter Market 2019: Global penetration.\label{globalsmartmeter}}  
\end{center} 
\end{figure*} 

\begin{figure*}[b!]
\begin{center}  
\includegraphics[width=0.835\textwidth]{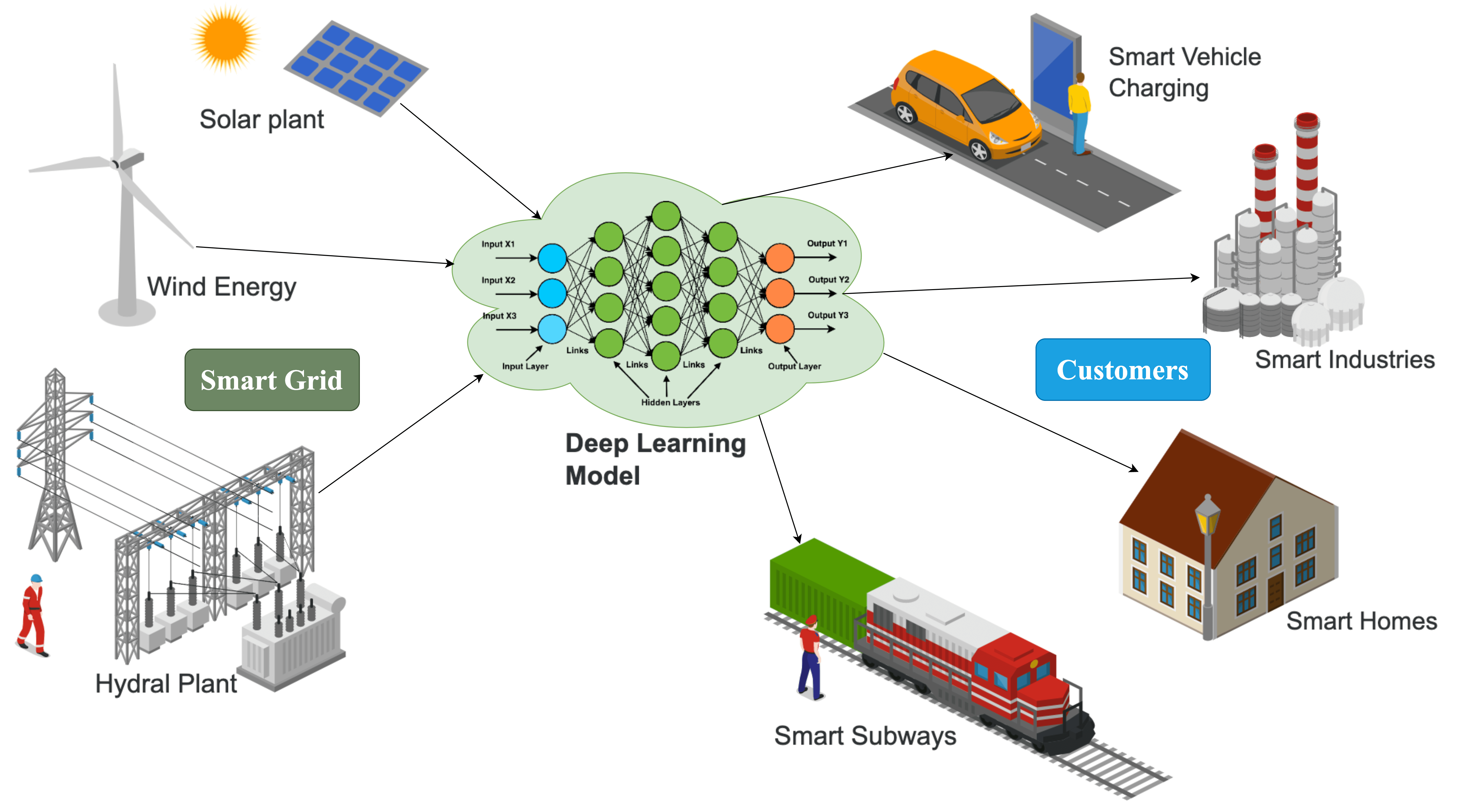}  
\caption{DL-based power transmission from Smart grid towards customers.\label{smartgrid}}  
\end{center} 
\end{figure*}

Smart meters dispose of embedded communication possibilities where several solutions are used, including power line communications (PLC), wireless low-power wide-area network (LPWAN) standards, and other wireless options, including RF Mesh (Radio Frequency mesh-based systems \cite{chakraborty2016advanced,andreadou2016telecommunication,varsier2017capacity,li2017research}.  According to recent research, smart meters will be the second most significant vertical IoT application in 2023 as deployed by energy and water utilities and have the potential to contribute over one-third of the global LPWA device connections (see Fig. \ref{globalsmartmeter}).

Mainly LoRa, Sigfox, and NB-IoT seem to be promising wireless technologies to enable smart metering. Following are some of the common examples of smart metering:
\begin{itemize}
\item In Sweden, Telia was able to show that the TCO of NB-IoT was less than that of PLC and RF Mesh for last-mile connectivity. Along with its system integration partners, Telia is converting and managing more than 2 million of the 5.4 million electric meters across the country with cellular IoT.
\item In France, LoRa is being used by Nova Veolia and its subsidiary Birdz, in collaboration with Orange Business Services, to connect over 3 million water meters to the LoRa network of Orange. Their goal is also to read more than 70\% of their meters remotely by 2027.
\item In Japan, 850,000 NICIGAS (Nippon Gas) gas meters across the country will get a smart makeover by the end of 2020, thanks to a retrofitted gas meter reader developed by UnaBiz and SORACOM. The gas consumption data are transmitted to NICIGAS' IoT data platform, via Sigfox's Japan-wide 0G wireless network.
\end{itemize}

The rapid development of new wireless technologies for IIoT is impacting the Europe's smart metering market \cite{nilsson2018smart,van2019smart,draetta2020emerging,uribe2016state}. For instance, Berg Insight states, adding that DSOs are looking for new projects on smart grid and roll-outs in 2020 with a wide range of various wireless technologies with many advantages compared to PLC technologies \cite{masum2019iot,weiss2016digital}. 

In the future smart cities, the smart grid will be the primary energy provider that involves energy-efficient resources (like renewable energy), smart meters, and other intelligent applications  \cite{kabalci2016survey,xu2018survey,gungor2011smart}. One of the differences between smart energy grids and traditional grids is that in the traditional electric grid system, consumers are billed once a month on provided electrical resources \cite{li2018deep2018,he2017real,wang2013energy}. However, due to the dynamic requirements of energy resources, we need a smart device that uses both way communication of customers and suppliers, which is the basic concept of the smart grid.

In smart grids, the electricity is provided from the microgrid (a distributed energy provider company located at the local level) to the users \cite{kukuvca2016smart,faheem2018smart,palensky2011demand}. The telecom companies are making contracts with local authorities to provide an advanced metering infrastructure (AMI) between the smart meters and service providers, as we know that the concept of the smart grid is not only restricted to electric suppliers \cite{zhang2018review2018,vermesan2013internet}. In AMI, the service provider tracks electricity consumption in real-time and provides suggestions and feedback to the users on electricity consumption and requirement.

Due to the massive amount of data generated by smart grids and smart meters, data management, and processing for controlling smart pricing is an issue \cite{li2019hybrid, parra2019implementation}. Hence, it is critical to have well-defined communication infrastructures and requirements for service providers with the customers through which power outage will be predictable and minimized (see Fig. \ref{smartgrid}). This will help industries in a way that they change their operation timing to avoid unscheduled power outage. 

\begin{figure*}[b!]
\begin{center}  
\includegraphics[width=0.9\textwidth]{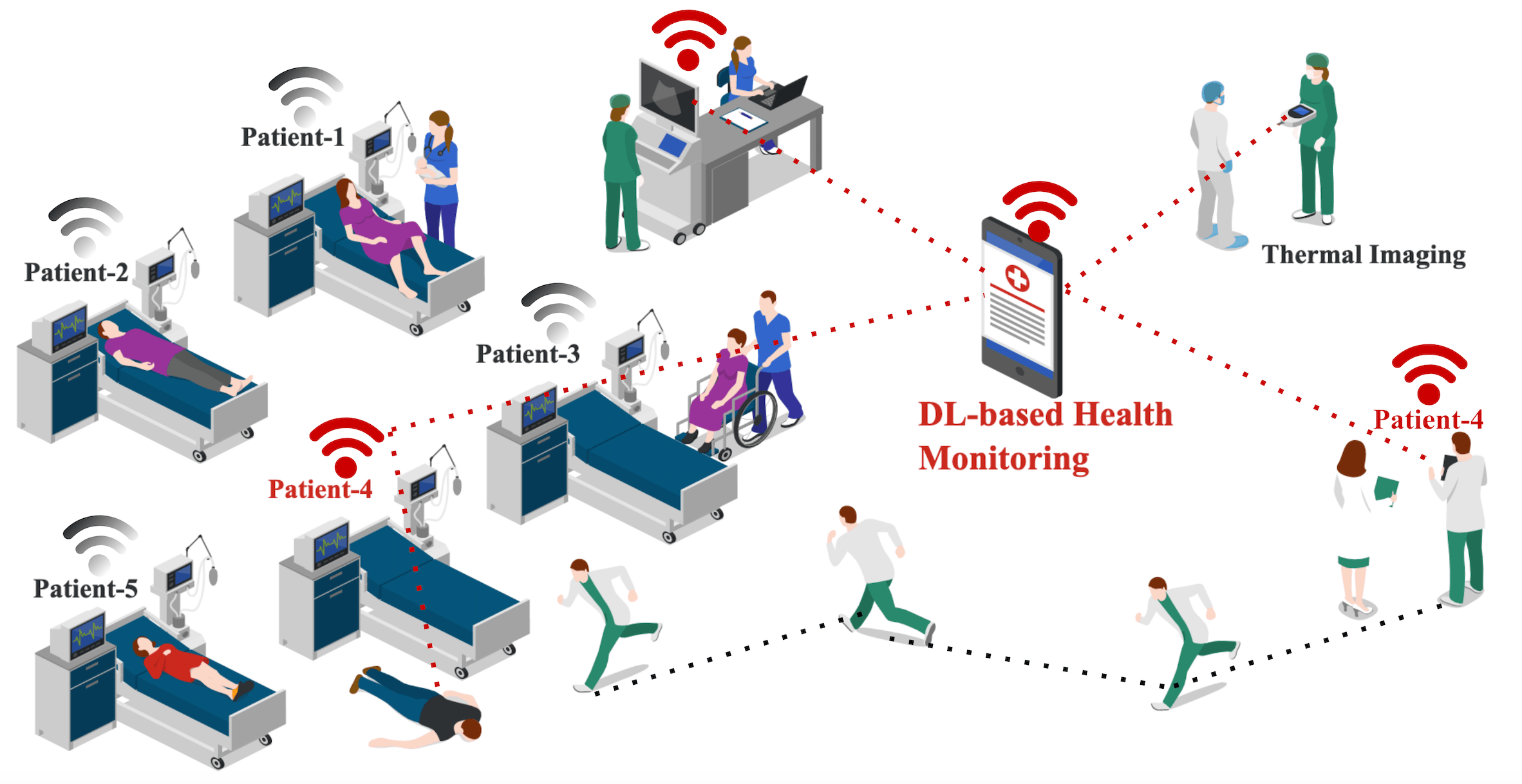}  
\caption{DL-based remote healthcare monitoring.\label{healthcare}}  
\end{center} 
\end{figure*} 

The data and communication will differ in frequency and type based on the users. For instance, a smart home will be less communicative as compared to a smart factory (which will be high communicative due to shifting in there power requirements between day and night and seasons \cite{wang2018deep3,esmalifalak2014detecting}.) Therefore, the smart grid team needs sophisticated data communication and flow management technologies. In the case of an industrial environment, different topological configurations and settings would be required for different flows, also illustrated in Fig. \ref{smartgrid}. For example,  one flow will be from the service provider to smart meters, and the other will be inside the factory from the smart meter to the different types of machinery and utilities. In such dynamic scenarios, a smart datacenter (having DL) with SDN and NFV functionalities would be needed for data management \cite{yan2018multi}.

\subsection{DL in Remote Healthcare Monitoring}
Healthcare is one of the most important and necessary industries, which offers millions of people value-based healthcare services around the world \cite{liang2014deep,min2017deep}. Therefore,  it is the world's top revenue generated business with a revenue of almost \$1.668 trillion in the United States alone. The value proposition of this industry is quality, value, and outcome. Therefore, it requires innovation in this industry, in which even stakeholders and specialists are investing highly in delivering the value prepositions of healthcare \cite{islam2015internet,tokognon2017structural}. Smart healthcare by technology is no longer a flight of fancy but is a requirement of health care due to the increasing global population.

The digital technologies are playing a critical role in various aspects of smart healthcare systems, such as patient care, billing, and handling medical records, and developing alternate staffing models. The research on DL in healthcare is seeking a gradual acceptance \cite{zhao2016deep,tuli2020healthfog,saba2019present}. For example, Google recently introduced a DL technique that can detect cancerous tumors. Moreover, researchers at Stanford University are investigation DL for the detection of skin cancer. In this context, DL is lending a hand in diverse healthcare situations, analyzing massive data, suggesting outcomes, and providing timely risk scores (see Fig. \ref{healthcare}). In the following, we provide some of the top utilization of DL in the healthcare industry:

\begin{itemize}
\item \textbf{Diagnosis:} One of the most significant uses in healthcare for DL is the diagnosis of a certain disease, including genetic disease and cancers at initial stages (which is hard to identify) \cite{tamilselvan2013failure,tuli2020healthfog}. The most common approach would be using DL models of identifying disease using classification or detection models trained on using supervisor learning. 

\item \textbf{Drug discovery and manufacturing:} AI and DL based algorithms help in the early-stage drug discovery process as a primary clinical application \cite{lavecchia2019deep,preuer2019interpretable,klambauer2019machine}. Drug discovery also lies in the R\&D section working next-generation precision and sequencing of medicines. And this is useful in finding the new and alternate ways of therapy for the multifactorial disease. The DL technique currently uses unsupervised learning for prediction; it only identifies patterns. For example, many researchers are using GAN network for finding vaccines or drugs for Corona disease by using GNOME \cite{lindelof2019deep,rifaioglu2019recent,maragakis2020deep}.

\item \textbf{Medical Imaging Diagnosis:} The recent breakthroughs in DL algorithms lead to the advancement of computer vision tools in the medical sector \cite{lundervold2019overview,biswas2019state,fu2019machine}. One such example is the diagnosis using medical imagery. For instance, the project by Microsoft named InnerEye, that utilizes image analysis as a diagnostic tool. Medical imagery analysis has may variables that arise at any moment and are discrete in nature \cite{sahiner2019deep,guo2019deep}. Therefore, the modeling of such variables using a complex mathematical analysis is quite difficult. With the use of DL algorithms, a ceratin model is easy to learn using data samples. Accordingly, the diagnosis approach based on DL will result in the classification of images into different categories like normal, abnormal, etc. \cite{currie2019machine,haskins2020deep,chan2020deep}.

\item \textbf{Smart Health Records:} DL and data analytic can also be useful in facilitating smart health care records \cite{simsek2020deep,xue2019explainable} as the maintaining of health care records is exhausting process. Even by using technology like data entry, it takes a lot of time to do so \cite{harerimana2019deep,amin2019cognitive}.

\item \textbf{Clinical Trial and Research:} Generally, clinical trials and research takes quite long time and sometimes can even take years to complete.  DL can enable prospective applications in clinical trials by reducing both cost and time for the Pharma industry \cite{stead2018clinical,wang2016perspective,shen2019systematic}. Applying DL-based data analytics help researchers to find potential clinical trials among wide variety of data points, such as social media use, previous doctor visits, etc. \cite{liang2018virtual,dwivedi2019artificial,wang2019deep}. This way, both time and resources for the clinical trials are minimized.

\item \textbf{Outbreak Prediction:} 
Today, the data scientists can predict outbreaks (malaria or severe chronic infection) using a large amount of data fed to Neural Networks. The data is collected from different open data platforms (Kaggle, etc.) or by themselves from real-time social media posts and updates, satellites, and websites information \cite{ardabili2020covid,ayyoubzadeh2020predicting,sharma2015malaria,zhang2017comparative,alessa2019preliminary}. 
Thanks to the breakthrough of AI and DL in predictions and the availability of high amounts of data in today's life, it is easy to predict and monitor epidemics around the globe \cite{chen2017disease,lu2017motor,jia2019predicting}. The prediction of these outbreaks helps prepare for the counter activities, especially in low-income countries, where there is a lack of crucial medical infrastructure. ProMED-mail is an example of monitoring evolving diseases and provide information regarding new emerging outbreaks in real-time with an internet-based reporting platform.

\end{itemize}

\subsection{DL in Enhancing Human Resources}
DL can bring loTs of facilities for the human resource (HR) management in any industry \cite{tambe2019artificial,lu2020service,bogoviz2020perspective,arnold2016industrial}. 
The managers in any company prefer value beyond the numbers in understanding what is happening, i.e., whether their leaders and executive staff need extra information or help in a particular area to point them in the right direction \cite{tsui2000role,pera2019towards,dahlbom2019big}. Thanks to the DL based algorithms, which can predict employee attrition and enable DL to work in the HR department.

DL can react faster in finding insights and inferences as in the changing and evolving key performance indicators (KPIs) than people who take time and a ream of the workforce \cite{buettner2013cognitive,kiel2017influence}. For instance, in the job application process, most of the companies use application tracking and assessment algorithms and platforms (like LinkedIn), which help in automation and making the process faster when there is a high volume of applicants for a specific role \cite{liebowitz2001knowledge,hmoud2019will,dirican2015impacts,breivold2015internet}. There is a dire need of DL based platform which is capable of giving calibrated guidance without humans, help in allowing more people to grow their skills, career, and stay engaged. 

On such example of DL in HR is Google's People Analytics, which is a pioneer in building engineers for performance-management at the enterprise level. People Analytics helped solve the fundamental problem of employee-related to business, focused on improving the lifestyle of "Googlers", their productivity, and overall wellness. This project posed some interesting questions (like the ideal size for a team or department) but focused on finding new ways of using data to find answers to such questions. In Short, DL in HR helps pave the new way for more valuable programs in less time by illuminating the development of a more people-centric approach \cite{paschek2017organizational,jarrahi2018artificial}. It also reduces personal biases in programs, and the need for administration, by increasing individual development.

\subsection{DL in Mining Industry}
The IBM states that in a lifetime, each person needs around 3.11 million pounds of mineral, fuels, and metal \cite{ibm_2019}. Therefore data analysis is a critical requirement for efficient use of the minerals \cite{lee2019case,ri2015defects,trappey2019patent}. Nevertheless, mining is a risky and expensive task to perform \cite{ali2020artificial}. Therefore, the use of IIoT will help improve production, avoid unnecessary waste, improve safety, and reduce costs for the mining industry. For example, data collection and analytic before the digging process can help save both time and the corresponding cost.

Similarly, IIoT can improve the mining industry's safety by overcoming the risks of suffocation, rock sliding, and other similar scenarios  \cite{carvalho2017mining}. On top of IIoT, DL models can enable autonomous drilling and digging systems that can reduce the risks and improve efficiency \cite{hyder2019artificial,van2009virtual,phakathi2017introduction,chiong2019bio}. 

\subsection{DL in Agriculture Industry}
\begin{figure*}[h!]
\begin{center}  
\includegraphics[width=0.85\textwidth]{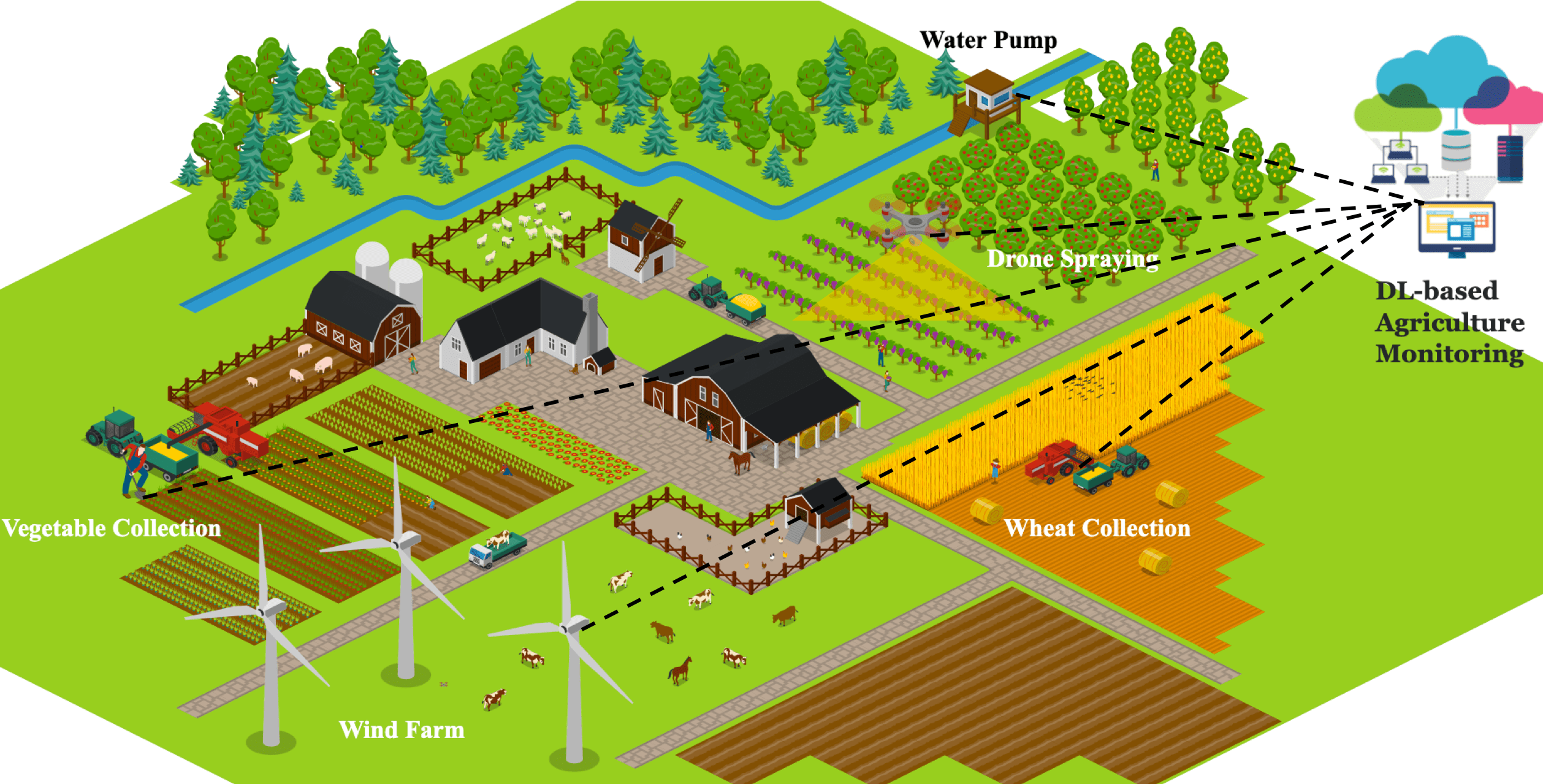}  
\caption{DL-based agriculture monitoring.\label{Fig13}}  
\end{center} 
\end{figure*}

DL-based IoT can play a fundamental role in the agriculture industry by making it smart and effective \cite{kamilaris2018deep,prathibha2017iot,suma2017iot,taylor2018climate,mekala2017survey,wang2018deepnew}. Recently, DL with IoT has become an essential technology for smart farming \cite{zhu2018deep,jin2020hybrid,park2018crops,kuwata2015estimating}.  For instance, drone technology DL assistance can enable effective ways of seeding, monitoring, and spraying crops. Moreover, DL-based IoT can also offer large scale irrigation systems. For example, a DL-based irrigation system with a single pump can offer an optimal solution, where the pump will operate only when there is a demand for more water. To find the specific area that needs more water, high-speed drones with wireless technology or moisture-based sensors deployed in the soil can provide the sensing information \cite{veeramani2018deepsort,coulibaly2019deep,ashqar2019identifying,lin2019research}. Once the area is identified, the pump will start operation, and the system will drive the water only when there is a need for water. Fig. \ref{Fig13} illustrates drone-based imagery to get the moisture-related information of specific agricultural land. Moreover, a trained model on specific datasets can be used to develop a water stress map, which can help find the location where water is needed.

\subsection{DL in Telecom Industry}
There is a variety of research studies regarding the use of DL in telecom industry \cite{prashanth2017high,ahmed2019transfer,khan2019customers}. DL techniques are used as tools for improving telecom networks' performance and helping in various compelling wireless technologies like massive MIMO, D2D, highly dense small cell networks, etc. \cite{alshathri2019cellular,mishra2017comparative,hebbalaguppe2017telecom}.  For example, in \cite{huang2017study}, a reinforcement learning method is investigated to optimize the scheduling of packet transmission in the cognitive-IoT networks. The main idea was to maximize the throughput's in a multi-channel environment of cognitive nodes in selecting appropriate values of network parameters (transmission power, scheduling, and spectrum action). Furthermore, in \cite{thantharate2019deepslice}, the authors proposed a DL-based model for classification and controlling of traffic in IoT networks. This model used a hybrid Recurrent Neural Networks with Convolution Neural Networks in predicting the class of packets by using features classification from packet headers. Another example is given in \cite{zahid2019big}, where the author studied different DL algorithms to analyze the challenges in wireless networks and their design optimization.

\subsection{DL in Transportation Industry}
The transportation industry is the backbone of any country where  ITS (intelligence transportation services) are becoming popular
\cite{zhang2011data,dimitrakopoulos2010intelligent,ozbay1999incident,hu2015vehicle,lv2014traffic,john2014traffic,wang2015deep}. 
In ITS, the roadside units can be equipped with a DL model to enable the internet of vehicle applications such as vehicle automation, supporting in-car entertainment, and providing location and context-aware services. 
Other applications pf DL in the transportation industry include smart parking and smart traffic light that control the traffic according to the situation rather than having some fixed rules \cite{hodges2019deep,bui2019big,wang2019enhancing,duan2014deep,huval2015empirical,chen2015road}. For example, in \cite{atallah2017deep}, the author used the D2D features of the LTE-A network for roadside units with a graphical neural network to enable smart transportation.

\subsection{DL in Waste Management Industry}
Making the process of waste management smart is one of the goals of the current waste management industry \cite{crooks1993giants,kabugo2020industry,sudha2016automatic,bircanouglu2018recyclenet}. There must be some flaws and inefficiencies in manual management, which is the cause of different diseases. The use of DL in waste management will not only improve the overall process, but it will also help us in speeding up the recycling process and utilizing the wastage effectively by linking the waste-collecting at the disposal to the recycling industry \cite{sousa2019automation,vafeiadis2019data,ajayi2016reducing,lapre2000behind}. The DL will properly allocate different resources following the quantity and type of specific waste like glass, paper, organic, etc. The DL will not only establish the connection between various industries like waste-to-energy and waste management authorities but will effectively manage the transfer and collection of waste material.

\subsection{DL in Advertisement Industry}
The advertisement industry in the world is moving toward smart advertisement by using DL algorithms. For example, YouTube, Facebook and other social media apps uses recommendation algorithms for their advertisements \cite{mcstay2016digital,lombard2001interactive,carter2002scheduling,li2019special}. In the same way, advertisements can be made smarter on large billboards screens updated by the trends in that location found by using the DL algorithms \cite{li2010exploitation,kietzmann2018artificial,cha2018artificial}. Consequently, at hyper-marts or shopping malls, advertisements on a small screen located on a specific location can be customized and changed from user-to-user by applying DL on the user's 
previous  data\cite{evans2009online,deng2019smart,lee2020digital}.

\section{Key Challenges in DL-IIoT}
We have discussed the importance of DL in various industries; however, its successful implementation in the industrial environment and obtaining trustful and useful results are not easy. It requires domain understanding with a problem-solving mindset and people with statistical analysis background who are good at finding valuable insights into the data. In the following, we discuss some of the key challenges of DL-assisted IIoT.

\subsection{Complexity}
Complexity is one of the major issues of DL models, which requires an extra effort to solve  \cite{li2018learning,meidan2018n}. One of the problems regarding DL performance is the time-consumption of the training phase and computation requirement due to the complexity of the model and large industrial dataset. The second issue is the lack of a large number of training samples in industrial scenarios, which reduces the accuracy and efficiency of models by overfitting. The problem of complexity can be tackled by using a tensor-train deep compression (TTDC) model for learning different features efficiently from the industrial data \cite{roopaei2017deep}. This technique compresses a large number of factors in the DL model, which further improves the model's speed. 

\subsection{Selection of Algorithm}
There are several widely available popular DL algorithms to be used in IIoT applications \cite{zhang2019machine}. These algorithms can work in any generic scenario, but selecting an algorithm for specific industrial applications should be based on particular guidelines available. For each algorithm, it is essential to know which DL algorithm will give its best in which scenario \cite{rymarczyk2019comparison}. The improper selection of a DL algorithm can cause many issues like the production of garbage outputs, which leads to a waste of time, effort, and money.

\subsection{Selection of Data}
A DL algorithm's success is directly correlated with the training data, which is well described by a famous saying, ``Garbage in, Garbage out." The correct type and amount of data are very critical for any DL algorithm \cite{wang2020novel}. It is vital to avoid such data that causes selective bias and make or select the data representative of the case in an industrial process.

\subsection{Data Preprocessing} 
After selecting the data, the next crucial step would be to convert the messy data containing missing values, outliers, and valueless entries to the form of data that could be understood by statistics and DL algorithms \cite{deutsch2017using,jiang2018deep,zhang2019deep11}. This step includes parsing, cleaning, and preprocessing data like converting the other form of data into numbers, scaling of features (for preventing their dominance over others), and removing or replacing missing entries.

\subsection{Data Labelling} 
As we know, supervised DL algorithms are the easiest and most appropriate algorithm these days in terms of implementation, training, and deployment in various IIoT applications. On the other hand, unsupervised DL algorithms are harder to implement, sometimes requiring several unsuccessful iterations and a very lengthy training process \cite{bi2010advances}. Nevertheless, data labeling for supervised DL algorithms is challenging and can't be outsourced for advanced and intensive tasks. For example, labeling of medical imaging on which a classification model could be trained for diagnosis process needs domain experts such as doctors. However, the issue is that specialized medical experts view this activity as a time-consuming process \cite{zhou2017computation}. Besides these critical challenges, many other challenges exist in the field of DL in a smart industrial environment, including managing model versions, data version, reproducing the models, etc. DL is continuously evolving, and its feature's learning capabilities are constantly changing \cite{mehdiyev2017time}. But incorporating the newer versions and features in the DL setup can sometimes be a nightmare if the team finds that the earlier dataset, models, and features are not appropriately documented.

\section{Future Research Directions in DL-IIoT}
For better expressing the overall requirements of DL-based IIoT, there are various aspects of DL that shall be rectified in the future to implicate the DL in smart industries conclusively. The different aspects of DL that need enhancement include intelligent algorithms with improved efficiency and supporting better platforms. Therefore, in the following, we provide a few essential research directions for  DL-based IIoT.

\subsection{Low-latency and Ultra-Reliability} 
As discussed earlier, that smart industrial setup requires several synchronized processes that require low latency and improved reliability to achieve the necessary performance \cite{heo2018super,yang2019learning,zuo2018low}. Moreover, the DL methods applied in IIoT should be able to handle these issues along with other parameters such as network deployment and resource management \cite{bennis2018ultrareliable}. Nonetheless, the competency and usefulness of DL-based IIoT scenarios are still in the evolving stage, exclusively demanding the strict low latency and ultra-reliability requirements in IIoT. Hence, research efforts are required in this direction to establish a theoretical and practical background for DL-IIoT to guarantee low-latency and ultra-reliable communication.

\subsection{DL-enabled Cloud/Edge Computing} 
Fast and efficient computing is another main feature that can affect not only the latency and reliability but many other performance parameters in smart industries \cite{ai2018edge}. As discussed, the IIoT requires powerful and useful tools to compute the big data obtained from various processes and analyze it at specific platforms, including servers, transmission mediums, and storage devices \cite{hossain2016cloud,hatcher2018survey,chen2019distributed}. Presently, hybrid cloud-edge computing is used to perform fast and effective computations and offers comprehensible computing infrastructure for IIoT. Nevertheless, to deal with an adequate complex-learning issue, the training-in-device processing is not considered as a feasible option \cite{chen2014big,pan2017future,corcoran2016mobile}. The reason behind this is the limited storage and low processing power of the devices, and the overall complexity of the problem. Therefore, it is considered suitable to use the edge-based infrastructure of computing due to its capability to reduce latency and improve the learning process in the network.
Nevertheless, the integration of DL and edge-based computing infrastructure for IIoT is still an open research problem  \cite{li2017consortium}. Specifically, the combined realization of distributed and parallel learning for edge-based designs requires additional optimization to achieve higher productivity, self-organization, and lower runtime.

\subsection{Intelligent Sensing and Decision Making} 
The controlling issues in DL-based IIoT consist of both sensing and evaluation procedures with the massive number of actuators and sensors \cite{gungor2009industrial,wang2006condition}. This will enable smart sensing capabilities in IIoT, such as prediction, categorization, and decision, directly controlling the overall system. For instance, in a smart manufacturing environment, intelligent sensing and useful decision-making aptitudes are strict enough with no allowed chances of economic loss and safety problems caused by failure  \cite{popescu2015economic}. Therefore, the successful implementation of DL-IIoT has strict requirements of effective prediction, categorization, and decision that can only be achieved with intelligent sensing-and-decision.

\subsection{E-Learning and Re-learning} 
IIoT is capable enough to enable smart and highly scalable connected industrial systems, where the handling of computation and networking are performed in a fully dynamic environment \cite{zander2012self}. It is a matter of the fact that DL models necessitate for pre-training to provide precise outputs from the learning processes \cite{willems2018learning,moreno2009reconfigurable}.  Moreover, with the dynamic and evolving nature of the IIoT systems, DL techniques should adapt to the complexities of the industrial environment \cite{lu2019artificial}. One way is to enable adaptive DL techniques is to combine e-learning and re-training for updating DL models continuously; however, this requires further investigation.

\subsection{Distributed Deep Learning} 
For large-scale DL processes with massive training datasets and long training times, distributed DL is a dedicated technique that allocates various computation resources to work collaboratively on a single task \cite{luckow2016deep}. Distributed DL can perform multiple tasks such as data collection, data mining, and testing phases into the various number of distributed nodes that simultaneously work on it, and hence, solve the problem in a time-efficient manner. Therefore, distributed DL is considered one of the fundamental techniques to be implemented in the IIoT environment \cite{oyekanlu2018distributed}. However, the implementation of distributed DL in smart industries is not an easy task. The main challenge is identifying the way to manage the overall distributed computation resources.

\subsection{Light-weight Learning Platform} 
 IIoT consists of various connected, intelligent devices to make a smart industrial setup \cite{kastensson2014developing}. These devices include controllers, actuators, sensors that use different DL algorithms for learning and provide future predictions. In general, the computational capacity of such devices is limited, thus requiring light-weight learning platforms \cite{cantor2008automotive,tan2018embedded}. This way, the learning process for various industrial devices will improve, resulting in an intelligent IIoT network with low computational complexity and improved network lifetime. One such example is the use of hardware-in-the-loop (HIL) platform of simulation \cite{lehfuss2012comparison,fathy2006review,tran2018internal,alvarez2017real,huo2019hardware,huerta2016power,mai2017comprehensive,lima2019hardware}. The HIL integrates the computations carried out for various process with the internal hardware of the sensors and actuators. The HIL platform's critical potential is the utilization of real-time data produced by deployed hardware testbeds.

\section{Conclusion}
Recent advancements in the Industrial Internet of Things (IIoT) and Deep learning (DL) have made the industrial setup smarter for various applications. According to Industry 4.0 standardization, there is a huge potential for DL in IIoT. Nevertheless, smart industries are still facing many challenges. In this paper, we presented the prospects of DL in IIoT. First, we discussed potential DL algorithms for IIoT networks, including Convolutional Neural Networks (CNNs), Auto Encoder (AE), Recurrent Neural Network (RNN), Restricted Boltzmann Machine (RBM) and its variants.
We then extend the discussion towards different use cases of DL-based IIoT, including predictive maintenance, asset tracking, smart meter, and smart grid, remote-healthcare monitoring, mining industry, transportation, telecom, and agriculture. We also discussed some of the remarkable challenges faced by the proper implementation of DL-based IIoT that include the complexity and selection of specific DL algorithms, preprocessing, and data labeling. Finally, we listed various future research directions for DL-based IIoT, such as low-latency, ultra-reliability, cloud/edge computing, intelligent sensing-and-decision. We believe that this survey can help both the academics and practitioners in the field of DL and IIoT to get an insight into various DL algorithms and their potentials in a specific smart industry.

\bibliographystyle{IEEEtran}
\bibliography{references}

\end{document}